\documentclass[pre,
   superscriptaddress,
   twocolumn,
   eqsecnum,
   showpacs,
   preprintnumbers,
   byrevtex]{revtex4-1}
\usepackage{amsfonts}
\usepackage{amssymb}
\usepackage{amsmath}
\usepackage{color}
\usepackage{graphicx}
\usepackage{subfigure}
\usepackage[pdftex,colorlinks=true]{hyperref}
\newcommand{\um}{u_m}
\newcommand{\uI}{u^{(i)}_m}
\newcommand{\uII}{u^{(ii)}_m}
\newcommand{\uIII}{u^{(iii)}_m}
\newcommand{\uIV}{u^{(iv)}_m}
\newcommand{\uV}{u^{(v)}_m}
\newcommand{\uVI}{u^{(vi)}_m}
\newcommand{\uVII}{u^{(vii)}_m}
\newcommand{\uVIII}{u^{(viii)}_m}
\newcommand{\uIX}{u^{(ix)}_m}

\newcommand{\del}{\Delta x}

\begin{document}
%
%
%

\preprint{LA-UR 10-01022}

\title
   {Stability and dynamical properties of Rosenau-Hyman compactons \\
   using Pad\'e approximants}

\author{Bogdan Mihaila}
\email{bmihaila@lanl.gov}
\affiliation{Materials Science and Technology Division, 
Los Alamos National Laboratory, Los Alamos, New Mexico 87545, USA} 

\author{Andres Cardenas}
\email{andres.cardenas@nyu.edu}
\affiliation{Materials Science and Technology Division, 
Los Alamos National Laboratory, Los Alamos, New Mexico 87545, USA}   
\affiliation{Physics Department, New York University, New York, NY 10003, USA}
\affiliation{Mathematics Department, Cal Poly Pomona, Pomona, CA 91768, USA}

\author{Fred Cooper}
\email{cooper@santafe.edu}
\affiliation{Santa Fe Institute, Santa Fe, NM 87501, USA}
\affiliation{Theoretical Division and Center for Nonlinear Studies, 
Los Alamos National Laboratory, Los Alamos, New Mexico 87545, USA}

\author{Avadh Saxena}
\email{avadh@lanl.gov}
\affiliation{Theoretical Division and Center for Nonlinear Studies, 
Los Alamos National Laboratory, Los Alamos, New Mexico 87545, USA}
   
\begin{abstract}
   We present a systematic approach for calculating higher-order derivatives of smooth functions on a uniform grid  using Pad\'e approximants. We illustrate our findings by deriving higher-order approximations using traditional second-order finite-differences formulas as our starting point. We employ these schemes to study the stability and dynamical properties of $K(2,2)$ Rosenau-Hyman (RH) compactons including the collision of two compactons and resultant shock formation. Our approach uses a differencing scheme involving only nearest and next-to-nearest neighbors on a uniform spatial grid.  The partial differential equation for the compactons involves first, second and third partial derivatives in the spatial coordinate and we concentrate on  four different  fourth-order methods which differ in the possibility of increasing the degree of accuracy (or not) of one of the spatial derivatives to sixth order. A method designed to reduce roundoff errors was found to be the most accurate approximation in stability studies of single solitary waves, even though all derivates are accurate only to fourth order. Simulating compacton scattering requires the addition of fourth derivatives related to artificial viscosity. For those problems the different choices lead to different amounts of ``spurious" radiation and we compare the virtues of the different choices.
\end{abstract}


\maketitle

\section{Introduction}

Since their discovery by Rosenau and Hyman in 1993~\cite{RH93}, compactons have found diverse applications in physics in the analysis of patterns on liquid surfaces~\cite{LD98}, in approximations for thin viscous films~\cite{BP96}, ocean dynamics~\cite{GOSS98}, magma dynamics~\cite{SSW07,SWR08}, and medicine~\cite{KEOK06}. Compactons are also the object of study in brane cosmology ~\cite{brane} as well as mathematical physics~\cite{KG98,CDM98}, and the dynamics of nonlinear lattices~\cite{SMR98,C02,CM06,PKJK06} to model the dispersive coupling of a chain of oscillators~\cite{PKJK06,RP05,PR06,R00}. Multidimensional RH compactons have been discussed in~\cite{R06,RHS07}. Recently, compact structures also have been studied in the context of a Klein-Gordon model~\cite{RK08,RK10}. A recent review of nonlinear evolution equations with cosine/sine compacton solutions can be found in Ref.~\onlinecite{RV09}.

Compactons represent a class of traveling-wave solutions with compact support resulting from the balance of both nonlinearity and nonlinear dispersion. Compactons were discovered by Rosenau and Hyman (RH) in the process of studying the role played by nonlinear dispersion in pattern formation in liquid drops using a family of fully nonlinear Korteweg-de Vries (KdV) equations~\cite{RH93},
\begin{equation}
   u_t + (u^l)_x + (u^p)_{xxx} = 0
   \>,
   \label{eq:kmn}
\end{equation}
where $u \equiv u(x,t)$ is the wave amplitude, $x$ is the spatial coordinate and $t$ is time.

RH called these solitary waves compactons, and Eq.~\eqref{eq:kmn} is known as the $K(l,p)$ compacton equation. The RH compactons have the remarkable soliton property that after colliding with other compactons they reemerge with the same coherent shape. However, unlike the soliton collisions in an integrable system, the point where the compactons collide is marked by the creation of low-amplitude compacton-anticompacton pairs~\cite{RH93}.

The RH generalization of the KdV equation~\eqref{eq:kmn} is 
only derivable from a Lagrangian in the $K(l,1)$ case. 
Hence, in general, Eq.~\eqref{eq:kmn} does not exhibit the usual energy conservation law. Therefore, Cooper, Shepard and Sodano~\cite{CSS93} proposed a different generalization of the KdV equation based on the first-order Lagrangian
\begin{equation}
   L(r,s) = \int \Bigl [ \frac{1}{2} \phi_x \phi_t - \frac{(\phi_x)^r}{r(r-1)}
   + \alpha (\phi_x)^s (\phi_{xx})^2 \Bigr ] \, dx
   \>.
   \label{eq:Llp}
\end{equation}
We note that the set $(l,p)$ in Eq.~\eqref{eq:kmn} corresponds to the set $(r-1,s+1)$ in Eq.~\eqref{eq:Llp}.
Since then, various other Lagrangian generalizations of the KdV equation have been considered~\cite{AC93,CHK01,DK98,CKS06,BCKMS09}. With the exception of Ref.~\cite{CHK01}, the structural stability of the resulting compactons was studied solely using analytical techniques such as linear stability analysis~\cite{AC93}, and an exhaustive numerical study of the stability and dynamical properties of these compacton solitary waves is needed.

In general, the numerical analysis of compactons is a difficult numerical problem because compactons have at most a finite number of continuous derivatives at their edges. Unlike the compactons derived from the Lagrangian~\eqref{eq:Llp}, the RH compactons have been the object of intense numerical study using pseudospectral methods~\cite{RH93,RHS07}, finite-element methods based on cubic B-splines~\cite{dF95,SC81} and on piecewise polynomials discontinuous at the finite element interfaces~\cite{LSY04}, finite-differences methods~\cite{IT98,RV07a,RV07b,RV08,RV09}, methods of lines with adaptive mesh refinement~\cite{SWSZ04,SWZ05}, and particle methods based on the dispersive-velocity method~\cite{CL01}.

Both the pseudospectral and finite-differences methods require artificial dissipation (hyperviscosity) to simulate interacting compactons without appreciable spurious radiation. The RH pseudospectral methods use a discrete Fourier transform and incorporate the hyperviscosity using high-pass filters based on second spatial derivatives. Using this approach RH showed successfully that compactons collide without any apparent radiation. However, because the pseudospectral methods explicitly damp the high-frequency modes in order to alleviate the negative effects due to the high-frequency dispersive errors introduced by the lack of smoothness at the edges of the compacton, the pseudospectral approach is not suitable for the study of high-frequency phenomena, and the usability of filters themselves has been called in the question~\cite{dF95}. In turn, finite-differences methods usually incorporate the artificial dissipation via a fourth spatial derivative term. However, in the absence of high-frequency filtering, these methods are marred by the appearance of spurious radiation~\cite{RV07a}. This radiation propagates both backward and forward and has an amplitude smaller by a few orders of magnitude than the compacton amplitude. Its numerical origin can be identified by a grid refinement technique.

Recently, Rus and Villatoro~\cite{RV07a,RV07b,RV08,RV09} introduced a discretization procedure for uniform spatial grids based on a Pad\'e approximant-like~\cite{baker} improvement of finite-differences methods. As special cases, this approach can be used to obtain the familiar second-order finite-differences methods and the fourth-order Petrov-Galerkin finite-element method developed by Sanz-Serna and co-authors~\cite{dF95,SC81}. Given the involved character of the Petrov-Galerkin approach based on linear interpolants described in Ref.~\cite{SC81}, we believe the work by Rus and Villatoro (RV) lends itself to further scrutiny.

In this paper we present a systematic derivation of the Pad\'e approximants~\cite{baker} intended to calculate derivatives of smooth functions on a uniform grid  by deriving higher-order approximations using traditional finite-differences formulas. Our derivation recovers as special cases the Pad\'e approximants first introduced by Rus and Villatoro~\cite{RV07a,RV07b,RV08,RV09}. We illustrate our approach for the particular case when second-order finite-differences formulas are used as the starting point to derive \emph{at least} fourth-order accurate approximations of the first three derivatives of a smooth function. This approach is equivalent to deriving the best differencing schemes involving only nearest and next-to-nearest neighbors on a uniform grid. We apply these approximation schemes to the study of stability and dynamical properties of  $K(p,p)$  Rosenau-Hyman compactons. This study is intended to establish the baseline for future studies of the stability and dynamical properties of $ L(r,s)$ compactons, which feature higher-order nonlinearities and terms with mixed-derivatives that are not present in the $K(p,p)$ equations. Hence, the numerical analysis of the properties of the $L(r,s)$ compactons of Eq.~\eqref{eq:Llp} is expected to be considerably more difficult. 

This paper is outlined as follows.  In Sec.~\ref{sec:pade}, we show that the approximation schemes discussed by RV~\cite{RV07a,RV07b} can be identified as special cases of a systematic improvement scheme that uses Pad\'e approximants to derive at least fourth-order accurate approximations for the first three spatial derivatives
$\uI$, $\uII$, and $\uIII$ 
(we have introduced the spatial discretization $x_m=mh$, $u(x) \rightarrow u_m$ and the roman numeral superscript denotes the order of spatial derivative at
$x_m$) 
by starting with  second-order finite-differences approximations. 
In general, one can begin with finite-differences approximations of any arbitrary even order, and improve upon these by at least two orders of accuracy by using suitable Pad\'e approximants. In Sec.~\ref{sec:approx} we discuss several special cases: Three of these cases describe approximation schemes that mix fourth-order accurate approximations for two of the derivatives $\uI$, $\uII$, and $\uIII$ with a
sixth-order accurate approximation for the third one. We also discuss the case of the ``optimal'' fourth-order approximation scheme. In the latter, all three derivatives are fourth-order accurate, but all the coefficients entering the Pad\'e approximants have values that result in a reduction of decimal roundoff errors. In Sec.~\ref{sec:res} we apply the above four approximation schemes to study the stability and dynamical properties of $K(2,2)$  Rosenau-Hyman compactons. We conclude by summarizing our main results in Sec.~\ref{sec:concl}.

%
%

\section{Pad\'e Approximants}
\label{sec:pade}

To begin, we consider a smooth function $u(x)$, defined on the interval $x \in [0,L]$, and discretized on a uniform grid, $x_m = m \, h$, with $m=0,1,\cdots,M$, and $h=L/M$. Pad\'e approximants of order $k$ of the derivatives of $u(x)$ are defined as \emph{rational} approximations of the form
\begin{align}
   \uI & \ = \frac{\mathcal{A}(E)}{\mathcal{F}(E)} \ \um + \mathcal{O}(\del^k)
   \>,
   \\
   \uII & \ = \frac{\mathcal{B}(E)}{\mathcal{F}(E)} \ \um + \mathcal{O}(\del^k)
   \>,
   \\
   \uIII & \ = \frac{\mathcal{C}(E)}{\mathcal{F}(E)} \ \um + \mathcal{O}(\del^k)
   \label{eq:c1}
   \>,
   \\
   \uIV & \ = \frac{\mathcal{D}(E)}{\mathcal{F}(E)} \ \um + \mathcal{O}(\del^k)
   \>,
\end{align}
where we have introduced the \emph{shift} operator, $E$, as
\begin{align}
   E^k \, \um = u_{m+k} \>.
\end{align}

In this language, the \emph{second}-order accurate approximation of derivatives based on finite-differences correspond to the Pad\'e approximants given by~\cite{RV07b}
\begin{align}
   \mathcal{A}_1(E)
   & \ = \frac{1}{2\del} \Bigl [ E - E^{-1} \Bigr ]
   \>,
   \\
   \mathcal{B}_1(E)
   & \ = \frac{1}{\del^2} \Bigl [ E - 2 + E^{-1} \Bigr ]
   \>,
   \\
   \mathcal{C}_1(E)
   & \ = \frac{1}{2\del^3} \Bigl [ E^2 - 2 E + 2 E^{-1} - E^{-2} \Bigr ]
   \>,
   \\
   \mathcal{D}_1(E)
   & \ = \frac{1}{\del^4} \Bigl [ E^2 - 4 E + 6 - 4 E^{-1} + E^{-2} \Bigr ]
   \>,
\end{align}
and $\mathcal{F}_1(E) = 1$. We note that \emph{even}- and \emph{odd}-order derivatives require approximants that are \emph{symmetric} and \emph{antisymmetric} in~$E$, respectively.

We also note that although all four operators, $\mathcal{A}_1(E)$, $\mathcal{B}_1(E)$, $\mathcal{C}_1(E)$, and $\mathcal{D}_1(E)$, lead to second-order accurate numerical approximations, the derivatives $\uIII$ and $\uIV$ involve the subset of grid points $\{x_m, x_{m\pm1}, x_{m\pm2} \}$, whereas the derivatives $\uI$ and $\uII$ involve only the subset of grid points $\{x_m, x_{m\pm1} \}$. Therefore, it is possible to design a numerical scheme that improves the order of approximation of the derivatives $\uI$ and $\uII$ by incorporating the additional grid points, $\{x_{m\pm2} \}$ (see Appendix~\ref{app:A}).

It is more challenging, however, to find a consistent approach that improves the order of approximation of all four lowest-order derivatives without extending the set of grid points. We will show next that the Pad\'e-approximant approach described here, allows us to provide a consistent approach involving only the grid points $\{x_m, x_{m\pm1}, x_{m\pm2} \}$ that includes three of these four derivatives.

In the following we will use extensively the Taylor expansion of $u(x)$ around $x_m$, i.e.
\begin{align}
\label{eq:taylor}
   &
   u_{m + k}
   \equiv
   u(x_m + k \del)
   =
   \um + \uI (k \del)
   \\ \notag & \
   + \uII \frac{k^2 \del^2}{2}
   + \uIII \frac{k^3 \del^3}{6}
   + \uIV \frac{k^4 \del^4}{24}
   \\ \notag & \
   + \uV \frac{k^5 \del^5}{120}
   + \uVI \frac{k^6 \del^6}{720}
   + \uVII \frac{k^7 \del^7}{5040}
   + \cdots
   \>.
\end{align}
The following two relationships follow immediately:
\begin{align}
\label{eq:sum}
   &
   \Bigl ( E^k + E^{-k} \Bigr ) \, u_m
   \equiv
   u_{m+k}
   +
   u_{m-k}
   =
   2 \, \um
   \\ \notag
   & \quad
   + \uII k^2 \del^2
   + \uIV \frac{k^4 \del^4}{12}
   + \uVI \frac{k^6 \del^6}{360}
   + \cdots
   \>,
\end{align}
and
\begin{align}
\label{eq:diff}
   &
   \Bigl ( E^k - E^{-k} \Bigr ) \, u_m
   \equiv
   u_{m+k}
   -
   u_{m-k}
   =
   2 \, \uI k \del
   \\ \notag
   & \quad
   + \uIII \frac{k^3 \del^3}{3}
   + \uV \frac{k^5 \del^5}{60}
   + \uVII \frac{k^7 \del^7}{2520}
   + \cdots
   \>.
\end{align}

To obtain a fourth-order accurate approximation of the derivatives, we can either begin by improving the third-order derivative, $\uIII$, or the fourth-order derivative, $\uIV$. Unfortunately, we cannot improve both these derivatives at the same time. Because in the compacton-dynamics problem~\cite{RH93,SC81,dF95,IT98,RV07a,RV07b,LSY04}, the fourth-order derivative enters only through the artificial viscosity term needed to handle shocks, we chose to improve the approximation corresponding to the third-order derivative, $\uIII$.

In the following we derive the operators $\mathcal{A}_2(E)$, $\mathcal{B}_2(E)$, $\mathcal{C}_2(E)$, and $\mathcal{D}_2(E)$, corresponding to the new fourth-order accurate Pad\'e approximants.

\subsection{Third-order derivatives}

Using Eqs.~\eqref{eq:c1} and~\eqref{eq:diff}, we obtain
\begin{align}
   \mathcal{C}_1& (E) \, \um
   \\ \notag & =
   \uIII
   + \uV \frac{\del^2}{4}
   + \uVII \frac{\del^4}{40}
   + \uIX\frac{17 \del^6}{12096}
   + \cdots
   \>,
\end{align}
or
\begin{align}
   \uIII
   =
   \mathcal{C}_1 & (E) \, \um
   - \uV \frac{\del^2}{4}
   + \mathcal{O}(\del^4)
   \>.
\end{align}
To eliminate the dependence on $\del^2$, we consider a linear combination of the second-order approximations of $\uIII$ on the same subset of grid points, $\{x_m, x_{m\pm1}, x_{m\pm2} \}$. This can be achieved by introducing an operator, $\mathcal{F}(E)$, symmetric in~$E$, such that
\begin{align}
   &
   \mathcal{F}(E) \, \uIII
   =
   \frac{1}{a} \,
   \Bigl [ \bigl ( E^2 + E^{-2} \bigr )
            + b \bigl ( E + E^{-1} \bigr )
            + c \Bigr ] \, \uIII
   \>,
\end{align}
such that
\begin{align}
   \mathcal{F}(E) \, \uIII
   =
   \mathcal{C}_1 (E) \, \um
   + \mathcal{O}(\del^k)
   \>.
\end{align}
Using Eq.~\eqref{eq:sum}, we obtain
\begin{align}
   &
   \mathcal{F}(E) \, \uIII
   =
   \frac{1}{a} \,
   \Bigl \{
   2 \uIII
   + 4 \uV \del^2
   + \uVII \frac{4 \del^4}{3}
   + \cdots
   \notag \\ & \
   + b \,
   \Bigl [
   2 \uIII
   + \uV \del^2
   + \uVII \frac{\del^4}{3}
   + \cdots
   \Bigr ]
   + c \, \uIII
   \Bigr \}
   \>.
\end{align}
Requiring that this approximation is \emph{fourth} order or better, we obtain the system of equations
\begin{align}
   a - 2 b - c & \ =
   2 \>,
   \\
   a - 4 b & \ =
   16 \>,
\end{align}
and its solution can be parameterized as:
\begin{align}
   a = 4\tau \>, \quad
   b = \tau-4 \>, \quad
   c = 2(\tau+3) \>.
\end{align}
It follows that we can write
\begin{align}
   {\mathcal{F}(E)} \ \uIII
   = & \
   \uIII
   + \uV \frac{\del^2}{4}
   + \uVII \Bigl ( \frac{1}{12} + \frac{1}{\tau} \Bigr ) \frac{\del^4}{4}
   \notag \\ &
   + \uIX \Bigl ( \frac{1}{60} + \frac{1}{\tau} \Bigr ) \frac{\del^6}{24}
   + \cdots
\label{eq:F}
   \>.
\end{align}
Hence, we have
\begin{align}
   \uIII = & \
   \frac{\mathcal{C}_1(E)}{\mathcal{F}(E)} \, \um
   + \uVII \Bigl ( \frac{1}{60} - \frac{1}{\tau} \Bigr ) \, \frac{\del^4}{4}
\label{eq:uIII}
   \\ \notag &
   + \uIX \Bigl ( \frac{43}{2520} - \frac{1}{\tau} \Bigr ) \, \frac{\del^6}{24}
   + \mathcal{O}(\del^8)
   \>.
\end{align}
For $\tau$ integer and $\tau \ge 5$, we obtain solutions with $a$, $b$, and $c$ positive integers.

\subsection{First-order derivatives}

Next, we calculate the corresponding Pad\'e approximants for the first-order derivative, $\uI$. We consider
\begin{align}
   \uI & \ = \frac{\mathcal{A}(E)}{\mathcal{F}(E)} \ \um + \mathcal{O}(\del^k)
   \>,
\end{align}
with $\mathcal{F}(E)$ given by~\eqref{eq:F}, and require that the order of the approximation is \emph{fourth} order or better. Therefore, $\mathcal{A}(E)$ must be an operator antisymmetric in~$E$. 

By definition, we introduce
\begin{align}
   &
   \mathcal{A}(E) \, \um
   =
   \frac{1}{\alpha \del} \,
   \Bigl [ \bigl ( E^2 - E^{-2} \bigr )
            + \beta \bigl ( E - E^{-1} \bigr )
   \Bigr ] \, \um
   \>,
\end{align}
and solve
\begin{align}
   \mathcal{F}(E) \, \uI
   =
   \mathcal{A} (E) \, \um
   + \mathcal{O}(\del^k)
   \>.
\end{align}
We have
\begin{align}
   &
   \mathcal{A}(E) \, \um
   =
   \frac{1}{\alpha} \,
   \Bigl \{
   4 \, \uI
   + \uIII \frac{8 \del^2}{3}
   + \uV \frac{8 \del^4}{15}
   \notag \\ & \qquad
   + \uVII \frac{16 \del^6}{315}
   + \cdots
   + \beta \,
   \Bigl [
   2 \, \uI
   + \uIII \frac{\del^2}{3}
   \notag \\ & \qquad
   + \uV \frac{\del^4}{60}
   + \uVII \frac{\del^6}{2520}
   + \cdots
   \Bigr ]
   \Bigr \}
   \>.
\end{align}
To satisfy the requirement of a fourth-order accurate approximation for the first-order derivative $\uI$, we solve the system of equations
\begin{align}
   \alpha - 2 \beta = & \ 4
   \>,
   \\
   3 \alpha - 4 \beta = & \ 32
   \>,
\end{align}
and obtain the solution
\begin{align}
   \alpha = 24 \>, \quad
   \beta = 10 \>.
\end{align}
This gives
\begin{align}
   \mathcal{A}_2(E) \, \um
   = & \
   \uI
   + \uIII \frac{\del^2}{4}
   \\ \notag &
   + \uV \frac{7}{240} \, \del^4
   + \uVII \frac{23}{10080} \, \del^6
   \>,
\end{align}
and we can write
\begin{align}
   \uI = & \
   \frac{\mathcal{A}_2(E)}{\mathcal{F}(E)} \, \um
   \textcolor{black}{-} \uV \Bigl ( \frac{1}{30} - \frac{1}{\tau} \Bigr ) \, \frac{\del^4}{4}
\label{eq:uI}
   \\ \notag &
   \textcolor{black}{-} \uVII \Bigl ( \frac{1}{105} - \frac{1}{4\tau} \Bigr ) \, \frac{\del^6}{6}
   + \mathcal{O}(\del^8)
   \>,
\end{align}
with
\begin{align}
   \mathcal{A}_2(E) & \ =
   \frac{1}{24 \del} \,
   \Bigl [
      E^2 + 10 E - 10 E^{-1} - E^{-2}
   \Bigr ]
   \>.
\end{align}

\subsection{Second-order derivatives}

To calculate the corresponding Pad\'e approximants for the second-order derivative, $\uI$, we begin with
\begin{align}
   \uII & \ = \frac{\mathcal{B}(E)}{\mathcal{F}(E)} \ \um + \mathcal{O}(\del^k)
   \>,
\end{align}
where $\mathcal{F}(E)$ is given again by~\eqref{eq:F}, and require that the approximation is \emph{fourth}-order accurate or better. It follows that the operator $\mathcal{B}(E)$ must be symmetric in~$E$, e.g.
\begin{align}
   &
   \mathcal{B}(E) \, \um
   =
   \frac{1}{\alpha \del^2} \,
   \Bigl [ \bigl ( E^2 + E^{-2} \bigr )
            + \beta \bigl ( E + E^{-1} \bigr )
            + \gamma
   \Bigr ] \, \um
   \>,
\end{align}
and solve for
\begin{align}
   \mathcal{F}(E) \, \uII
   =
   \mathcal{B} (E) \, \um
   + \mathcal{O}(\del^k)
   \>.
\end{align}
We have
\begin{align}
   &
   \mathcal{B}(E) \, \um
   =
   \frac{1}{\alpha \del^2} \,
   \Bigl \{
   2 \um
   + 4 \uII \del^2
   + \uIV \frac{4 \del^4}{3}
   + \cdots
   \notag \\ & \
   + \beta \,
   \Bigl [
   2 \um
   + \uII \del^2
   + \uIV \frac{\del^4}{12}
   + \cdots
   \Bigr ]
   + \gamma \, \um
   \Bigr \}
   \>,
\end{align}
which gives the system of equations
\begin{align}
   2 \beta + \gamma = & \ - 2
   \>,
   \\
   \alpha - \beta = & \ 4
   \>,
   \\
   3 \alpha - \beta = & \ 16
   \>,
\end{align}
with the solution
\begin{align}
   \alpha = 6 \>, \quad
   \beta = 2 \>, \quad
   \gamma = - 6 \>.
\end{align}
Hence, we find
\begin{align}
   \mathcal{B}_2(E) \, \um
   = & \
   \uII
   + \uIV \frac{\del^2}{4}
   \\ \notag &
   + \uVI \frac{11}{360} \, \del^4
   + \uVIII \frac{43}{20160} \, \del^6
   + \cdots
   \>,
\end{align}
and we can write
\begin{align}
   \uII = & \
   \frac{\mathcal{B}_2(E)}{\mathcal{F}(E)} \, \um
   \textcolor{black}{-} \uVI \Bigl ( \frac{7}{180} - \frac{1}{\tau} \Bigr ) \, \frac{\del^4}{4}
\label{eq:uII}
   \\ \notag &
   \textcolor{black}{-} \uVIII \Bigl ( \frac{29}{840} - \frac{1}{\tau} \Bigr ) \, \frac{\del^6}{24}
   + \mathcal{O}(\del^8)
   \>,
\end{align}
with
\begin{align}
   \mathcal{B}_2(E) & \ =
   \frac{1}{6 \del^2} \,
   \Bigl [
      E^2 + 2 E - 6 + 2  E^{-1} + E^{-2}
   \Bigr ]
   \>.
\end{align}

\subsection{Fourth-order derivatives}

Because we chose to begin our derivation by improving the third-order derivative, $\uIII$, and both the finite-differences approximation for $\uIII$ and $\uIV$ already involve the entire subset, $\{x_m, x_{m\pm1}, x_{m\pm2} \}$, it follows that we are limited to a second-order accurate approximation for the fourth-order derivative, $\uIV$. The error corresponding to the Pad\'e approximant,
\begin{align}
   \uIV & \ = \frac{\mathcal{D}_1(E)}{\mathcal{F}(E)} \ \um + \mathcal{O}(\del^2)
   \>,
\end{align}
is obtained from the equation
\begin{align}
   \mathcal{F}(E) \, \uIV
   =
   \mathcal{D}_1(E) \, \um
   + \mathcal{O}(\del^2)
   \>.
\end{align}
Using $\mathcal{F}(E)$ from Eq.~\eqref{eq:F} and
\begin{align}
   &
   \mathcal{D}_1(E) \, \um
   =
   \uIV
   + \uVI \frac{\del^2}{6}
   + \uVIII \frac{\del^4}{80}
   + \cdots
   \>,
\end{align}
we find
\begin{align}
   \uIV & \ = \frac{\mathcal{D}_1(E)}{\mathcal{F}(E)} \ \um
   \textcolor{black}{+} \uVI \frac{\del^2}{12}
   +
   \mathcal{O}(\del^4)
   \>.
\end{align}

%
%

\begin{figure*}[t]
   \subfigure[\ (4,4,4) scheme]
      { \includegraphics[width=\columnwidth]{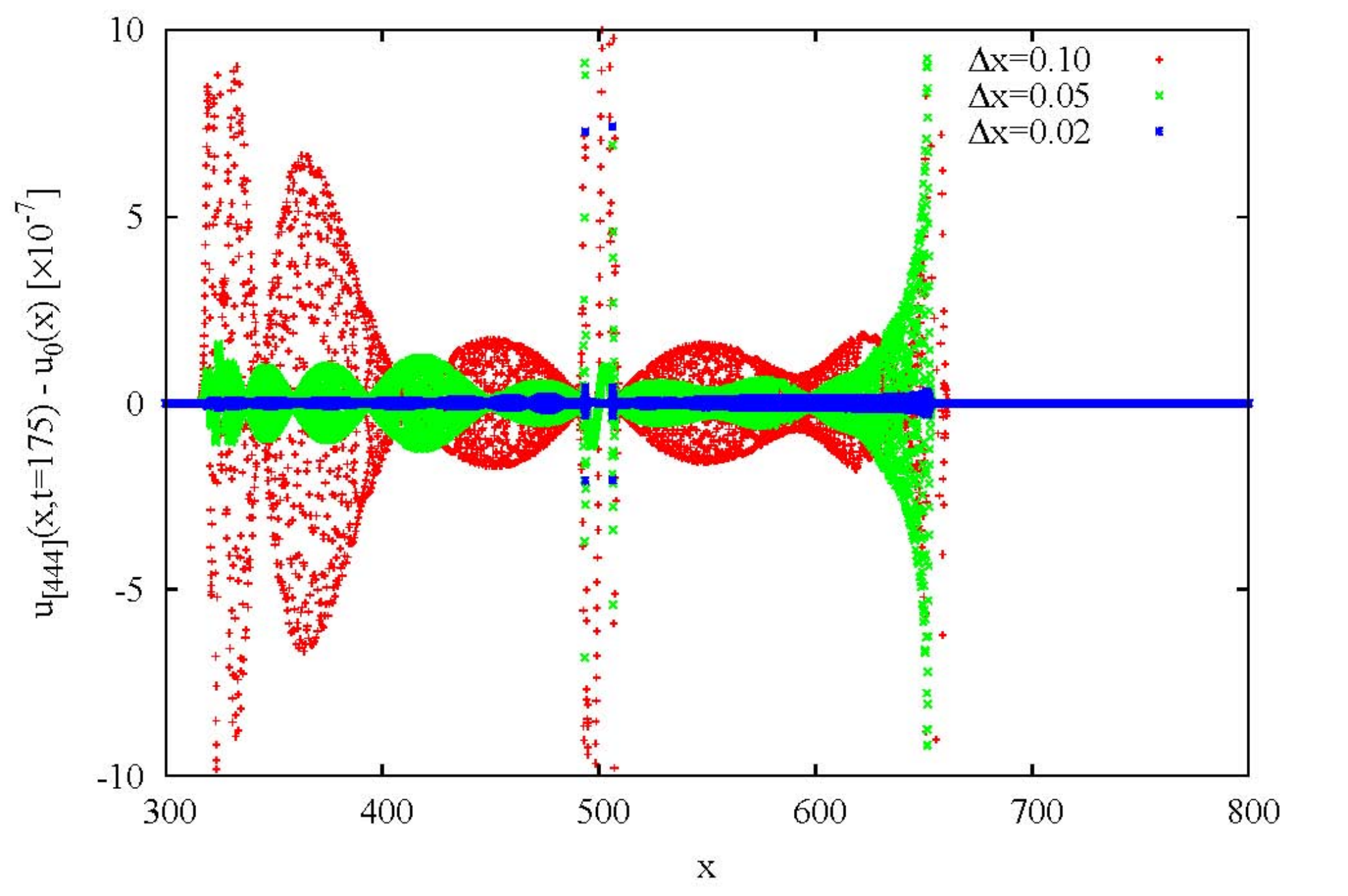} }
   \subfigure[\ (6,4,4) scheme]
      { \includegraphics[width=\columnwidth]{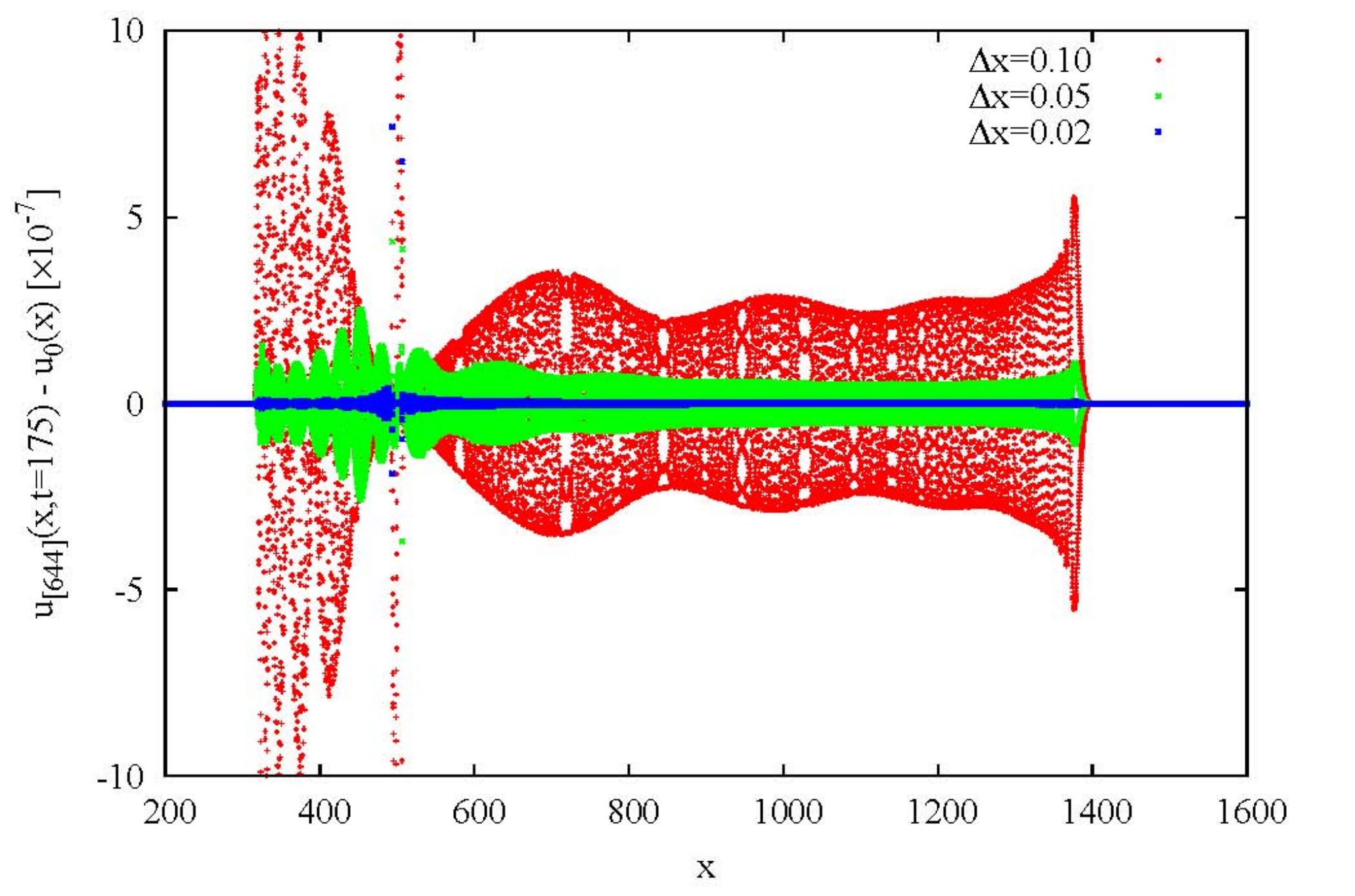} }
   \subfigure[\ (4,6,4) scheme]
      { \includegraphics[width=\columnwidth]{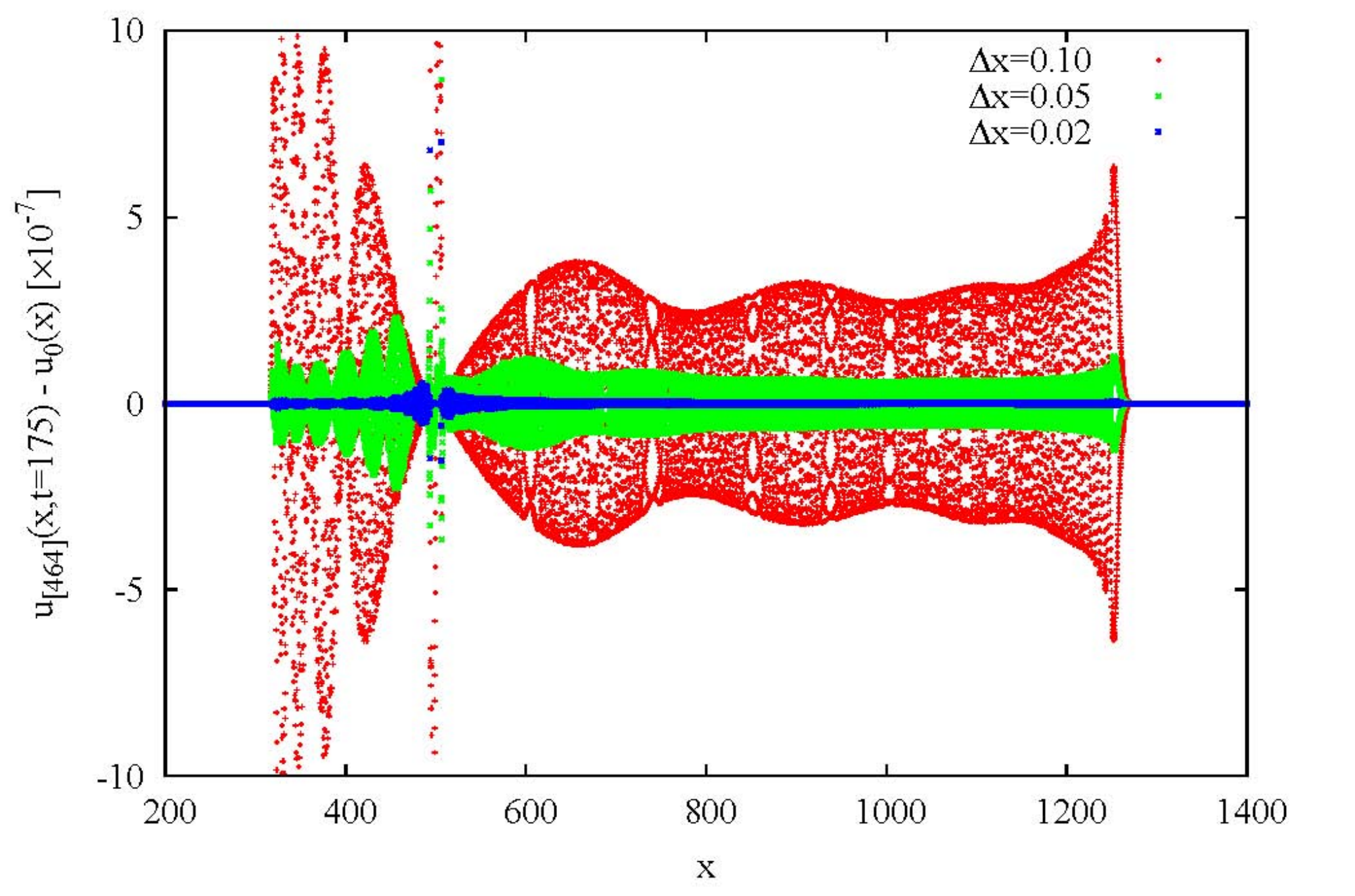} }
   \subfigure[\ (4,4,6) scheme]
      { \includegraphics[width=\columnwidth]{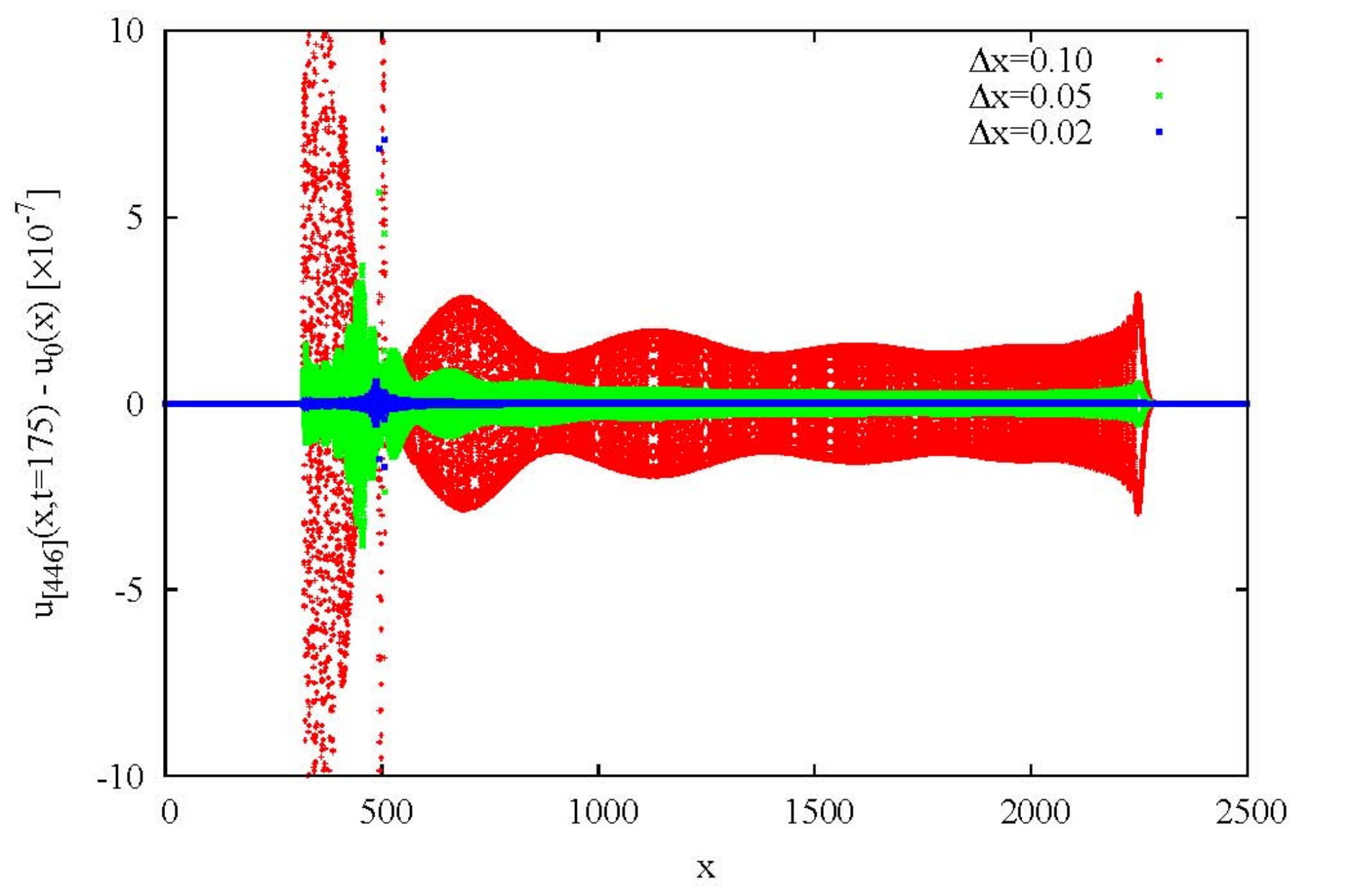} }
   \caption{\label{stab_1} (Color online) Study of the $K(2,2)$ (Rosenau-Hyman) compacton stability: For each numerical scheme we illustrate results at time $t$=175, the compacton propagation in its comoving frame with $\Delta t$=0.002 and $\Delta x$=0.1, 0.05, and 0.02. In all cases the radiation depicted here is a numerical artifact that is suppressed by reducing the grid spacing, $\Delta x$. This indicates that indeed the compacton is a stable solution of the $K(2,2)$  model.}
\end{figure*}

%
%

\section{Approximation schemes}
\label{sec:approx}

Based on the above considerations regarding Pad\'e approximants on the subset of grid points, $\{x_m, x_{m\pm1}, x_{m\pm2} \}$, it follows that we can always obtain a scheme that provides fourth-order accurate approximations for the derivatives $\uI$, $\uII$, and $\uIII$. It is however possible to obtain approximants that mix fourth-order accurate approximations for two of these derivatives with a sixth-order accurate Pad\'e approximant for the third one. We will discuss these special cases next, together with what may represent the ``optimal'' fourth-order approximation scheme.

{\rm{\textbf{(6,4,4)}} scheme:} This approximation scheme is an extension of the scheme introduced by Sanz-Serna \emph{et al.}~\cite{dF95,SC81}  using a fourth-order Petrov-Galerkin finite-element method, and corresponds to choosing $\tau=30$ in Eqs.~\eqref{eq:uI} and~\eqref{eq:uIII}. Then, we have
\begin{align}
   a = 120 \>, \quad
   b = 26 \>, \quad
   c = 66 \>,
\end{align}
and the coefficient of $\del^4$ vanishes in Eq.~\eqref{eq:uI}. Therefore, we obtain a sixth-order accurate approximation for the first-order derivative,
\begin{align}
   \uI =
   \frac{\mathcal{A}_2(E)}{\mathcal{F}_{[644]}(E)} \, \um
   \textcolor{black}{-} \uVII \frac{\del^6}{5040}
   + \mathcal{O}(\del^8)
   \>,
\end{align}
a fourth-order accurate approximation for the second-order derivative,
\begin{align}
   \uII = & \
   \frac{\mathcal{B}_2(E)}{\mathcal{F}_{[644]}(E)} \, \um
   \textcolor{black}{-} \uVI \frac{\del^4}{720}
   + \mathcal{O}(\del^6)
   \>,
\end{align}
and a fourth-order accurate approximation for the third-order derivative,
\begin{align}
   \uIII =
   \frac{\mathcal{C}_1(E)}{\mathcal{F}_{[644]}(E)} \, \um
   \textcolor{black}{+} \uVII \frac{\del^4}{240}
   + \mathcal{O}(\del^6)
   \>,
\end{align}
where we introduced the notation
\begin{align}
   \mathcal{F}_{[644]}(E) & \ =
   \frac{1}{120} \,
   \Bigl [
      E^2 + 26 E + 66 + 26 E^{-1} + E^{-2}
   \Bigr ]
   \>.
\end{align}

{\rm{\textbf{(4,6,4)}} scheme:} The coefficient of $\del^4$ in Eq.~\eqref{eq:uII} does not vanish for an integer value of $\tau$. To obtain a sixth-order accurate approximation for $\uII$, we require $\tau=180/7$. Then, we have
\begin{align}
   a = \frac{720}{7} \>, \quad
   b = \frac{152}{7} \>, \quad
   c = \frac{402}{7} \>,
\end{align}
and we obtain a fourth-order accurate approximation of the first-order derivative,
\begin{align}
   \uI =
   \frac{\mathcal{A}_2(E)}{\mathcal{F}_{[464]}(E)} \, \um
   \textcolor{black}{+} \uV \frac{\del^4}{720}
   + \mathcal{O}(\del^6)
   \>,
\end{align}
a sixth-order accurate approximation of the second-order derivative,
\begin{align}
   \uII = & \
   \frac{\mathcal{B}_2(E)}{\mathcal{F}_{[464]}(E)} \, \um
   \textcolor{black}{+} \uVIII \frac{11}{60480} \, \del^6
   + \mathcal{O}(\del^8)
   \>,
\end{align}
and a fourth-order accurate approximation of the third-order derivative,
\begin{align}
   \uIII =
   \frac{\mathcal{C}_1(E)}{\mathcal{F}_{[464]}(E)} \, \um
   \textcolor{black}{+} \uVII \frac{\del^4}{180}
   + \mathcal{O}(\del^6)
   \>,
\end{align}
where we introduced the notation
\begin{align}
   \mathcal{F}_{[464]}(E) & \ =
   \frac{1}{720} \,
   \Bigl [
      7 E^2 + 152 E + 402 + 152 E^{-1} + 7 E^{-2}
   \Bigr ]
   \>.
\end{align}

{\rm{\textbf{(4,4,6)}} scheme:} For $\tau=60$, the coefficient of $\del^4$ vanishes in Eq.~\eqref{eq:uIII} and we obtain a sixth-order accurate approximation for $\uIII$. We have
\begin{align}
   a = 240 \>, \quad
   b = 56 \>, \quad
   c = 126 \>,
\end{align}
and we obtain a fourth-order accurate approximation of the first-order derivative,
\begin{align}
   \uI =
   \frac{\mathcal{A}_2(E)}{\mathcal{F}_{[446]}(E)} \, \um
   \textcolor{black}{-} \uV \frac{\del^4}{240}
   + \mathcal{O}(\del^6)
   \>,
\end{align}
a fourth-order accurate approximation of the second-order derivative,
\begin{align}
   \uII = & \
   \frac{\mathcal{B}_2(E)}{\mathcal{F}_{[446]}(E)} \, \um
   \textcolor{black}{-} \uVI \frac{\del^4}{180}
   + \mathcal{O}(\del^6)
   \>,
\end{align}
and a sixth-order accurate approximation of the third-order derivative,
\begin{align}
   \uIII =
   \frac{\mathcal{C}_1(E)}{\mathcal{F}_{[446]}(E)} \, \um
   \textcolor{black}{-} \uIX \frac{\del^6}{60480}
   + \mathcal{O}(\del^8)
   \>,
\end{align}
where we have introduced the notation
\begin{align}
   \mathcal{F}_{[446]}(E) & \ =
   \frac{1}{240} \,
   \Bigl [
      E^2 + 56 E + 126 + 56 E^{-1} + E^{-2}
   \Bigr ]
   \>.
\end{align}
This scheme is an extension of the scheme introduced first by Rus and Villatoro~\cite{RV07b,RV07a}.

{\rm{\textbf{(4,4,4)}} scheme:} Finally, for the smallest value of $\tau$ leading to integer positive values of $a$, $b$, and $c$ (i.e. $\tau=5$), we obtain
\begin{align}
   a = 20 \>, \quad
   b = 1 \>, \quad
   c = 16 \>.
\end{align}
This gives a fourth-order accurate approximation of the first-order derivative,
\begin{align}
   \uI =
   \frac{\mathcal{A}_2(E)}{\mathcal{F}_{[444]}(E)} \, \um
   \textcolor{black}{+} \uV \frac{\del^4}{24}
   + \mathcal{O}(\del^6)
   \>,
\end{align}
a fourth-order accurate approximation of the second-order derivative,
\begin{align}
   \uII = & \
   \frac{\mathcal{B}_2(E)}{\mathcal{F}_{[444]}(E)} \, \um
   \textcolor{black}{-} \uVI \frac{29}{720} \, \del^4
   + \mathcal{O}(\del^6)
   \>,
\end{align}
and a fourth-order accurate approximation of the third-order derivative,
\begin{align}
   \uIII =
   \frac{\mathcal{C}_1(E)}{\mathcal{F}_{[444]}(E)} \, \um
   \textcolor{black}{+} \uVII \frac{11}{240} \del^4
   + \mathcal{O}(\del^6)
   \>,
\end{align}
where we have introduced the notation
\begin{align}
   \mathcal{F}_{[444]}(E) & \ =
   \frac{1}{20} \,
   \Bigl [
      E^2 + E + 16 + E^{-1} + E^{-2}
   \Bigr ]
   \>.
\end{align}

%
%

\section{Results}
\label{sec:res}

To compare the quality of the approximations discussed above, we specialize to the case of the $K(p,p)$ equation. In a frame of reference moving with velocity $c_0$, the $K(p,p)$ equation reads
\begin{equation}
   \label{Kpp}
   \frac{\partial u}{\partial t}
   - c_0 \, \frac{\partial u}{\partial x}
   + \frac{\partial u^p}{\partial x}
   + \frac{\partial^3 u^p}{\partial x^3}
   = 0, \qquad
   1< p \leq 3
   \>.
\end{equation}
For $p$ restricted to the interval $1 < p \leq 3$, the $K(p,p)$ equation allows for a compacton solution, with the simple form~\cite{IT98,RV07a,R98}
\begin{equation}
   u_c(x,t) = \alpha^\gamma \cos^{2\gamma} \Bigl [ \beta \xi(x,t) \Bigr ]
   \>,
   \qquad
   |\xi(x,t)| \leq \pi/(2\beta) \>,
\end{equation}
where $c$ is the compacton velocity and $x_0$ is the position of its maximum at $t = 0$, and we have introduced the notations $\xi(x,t) = x - x_0 - (c-c_0)t$, and
\begin{equation}
   \alpha = \frac{2cp}{p+1} \>,
   ~\beta = \frac{p-1}{2p} \>,
   ~\gamma = \frac{1}{p-1} \>.
\end{equation}

%
%

\begin{figure}[b!]
   \centering
   \includegraphics[width=0.9\columnwidth]{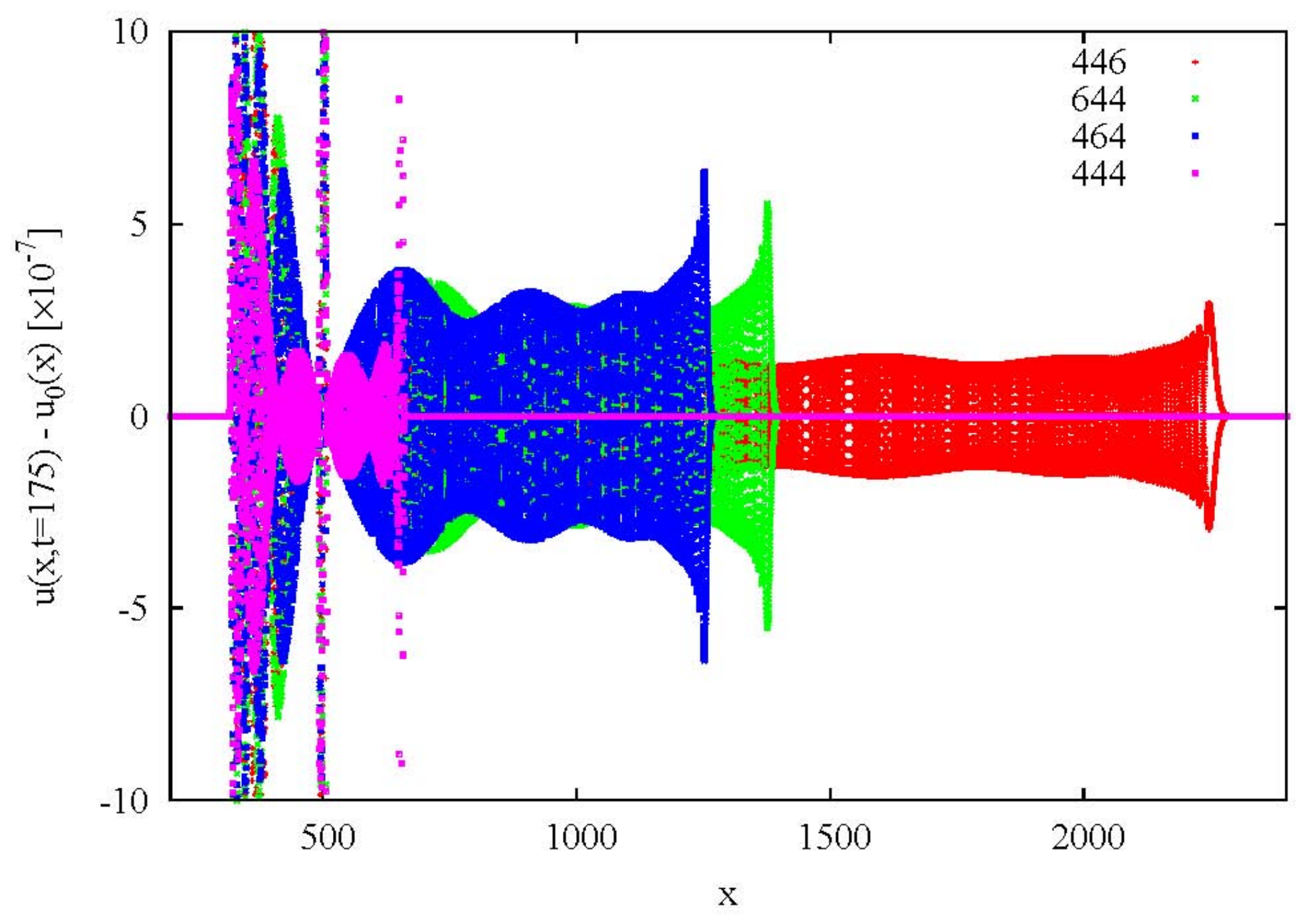}
   \caption{\label{stab_2}(Color online)
   Comparison of compacton-stability results as a function of numerical scheme. Results are shown at $t$=175, for compacton propagation in its comoving frame with $\Delta t$=0.002 and $\Delta x$=0.1.}
\end{figure}

%
%

Numerically, the lack  of smoothness at the edge of the compacton introduces numerical high-frequency dispersive errors into the calculation, which can destroy the accuracy of the
simulation unless they are explicitly damped (see e.g. discussion in Ref.~\cite{CHK01}). As such, we solve Eq.~\eqref{Kpp} in the presence of an artificial dissipation (hyperviscosity) term based on fourth spatial derivative, $\mu \, \partial^4 u / \partial x^4$, and we choose $\mu$ as small as possible to reduce these numerical artifacts while not significantly changing the solution to the compacton problem. We note nonetheless, that the addition of artificial dissipation results in the appearance of tails and compacton amplitude loss. 

Let us consider now the numerical solution of Eq.~\eqref{Kpp} by means of the fourth-order accurate Pad\'e approximants discussed here. In general, we can discretize Eq.~\eqref{Kpp} in space as
\begin{align}
   \mathcal{F}(E) \, \frac{\mathrm{d}u_m}{\mathrm{d}t}
   & 
   - \Bigl [ c_0 \mathcal{A}(E) - \mu \mathcal {D}(E) \Bigr ] u_m
   \notag \\ & 
   + \Bigl [ \mathcal{A}(E)  + \mathcal{C}(E) \Bigr ] (u_m)^p
   = 0
   \>.
\label{eq:1}
\end{align}
We consider a uniform grid in the interval $x\in[0,L]$ by introducing the grid points $x_m = m \Delta x$, with $m = 0,1,\cdots,M$ and the grid spacing $\Delta x = L/M$. In Eq~\eqref{eq:1}, 
we assume that $u_m(t)$ obeys periodic boundary conditions, $u_{M}(t) = u_0(t)$.

Following RV~\cite{RV09}, we have numerically discretized the time-dependent part of Eq~\eqref{eq:1} by implementing both the implicit trapezoidal (Euler) and the implicit midpoint rule in time. Correspondingly, we need to solve the following two approximate equations for Eq.~\eqref{eq:1}:
\begin{align}
   \mathcal{F}(E) \, & \frac{u_m^{n+1}-u_m^n}{\Delta t}
\label{eq:1trap}
   \\ \notag &
   - \Bigl [ c_0 \mathcal{A}(E) - \mu \mathcal {D}(E) \Bigr ] \frac{u_m^{n+1} + u_m^n}{2}
   \\ \notag & 
   + \Bigl [ \mathcal{A}(E)  + \mathcal{C}(E) \Bigr ] 
   \frac{(u_m^{n+1})^p + (u_m^n)^p}{2}
   \ = \ 0
   \>,
\end{align}
corresponding to the trapezoidal rule, and
\begin{align}
   \mathcal{F}(E) \, & \frac{u_m^{n+1}-u_m^n}{\Delta t}
\label{eq:1mid}
   \\ \notag &
   - \Bigl [ c_0 \mathcal{A}(E) - \mu \mathcal {D}(E) \Bigr ] \frac{u_m^{n+1} + u_m^n}{2}
   \\ \notag & 
   + \Bigl [ \mathcal{A}(E)  + \mathcal{C}(E) \Bigr ] 
   \Bigl ( \frac{u_m^{n+1} + u_m^n}{2} \Bigr )^p
   = 0	
   \>,
\end{align}
corresponding to the midpoint rule. Here we have introduced the notations, $u_m^n=u_m(t_n)$ and $u_m^{n+1} =u_m(t_n+\Delta t)$.

In the following, we further specialize to the case of the $K(2,2)$  equation ($p=2$), which allows for the exact compacton solution
\begin{equation}
   u_c(x,t) = \frac{4c}{3} \cos^2 \Bigl [ \frac{ x - (c-c_0)t }{4} \Bigr ]
   \>,
\label{eq:uc}
\end{equation}
in the interval $ |x - (c-c_0)t| \leq 2\pi$, where $c$ is the velocity of the compacton. We note that in our simulations pertaining the $K(2,2)$ compacton problem, we did not find any numerically-significant differences between stepping out the solution using the trapezoidal and the midpoint rules. This is consistent with the observation made by RV in Refs.~\cite{RV07a,RV07b}. Therefore, in the following we only present results obtained using the trapezoidal rule. Implementing both methods is however important for the purpose of future simulations of compactons exhibiting higher-order nonlinearities, e.g. in the case of the $L(r,s)$ compactons.

%
%

\begin{figure}[t!]
   \centering
   \includegraphics[width=0.9\columnwidth]{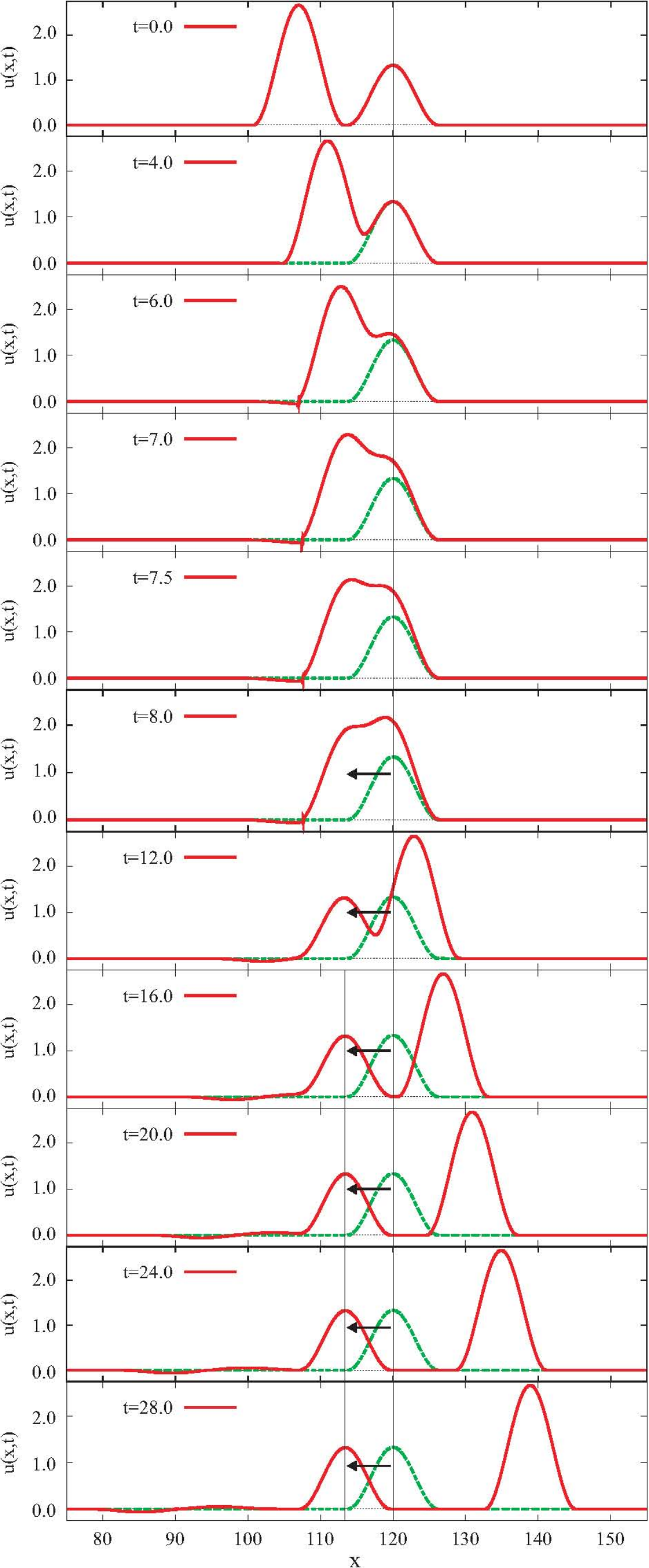}
   \caption{\label{coll_mov}(Color online)
   Collision of two compactons with $c_1=1$ and $c_2=2$. The simulation is performed in the comoving frame of reference of the first compacton, i.e. $c_0=c_1$, using the (6,4,4) scheme and a hyperviscosity, $\mu=10^{-4}$. The collision is shown to be inelastic, despite the fact that the compactons maintain their coherent shapes after the collision.}
\end{figure}

\begin{figure}[h!]
   \centering
   \includegraphics[width=0.9\columnwidth]{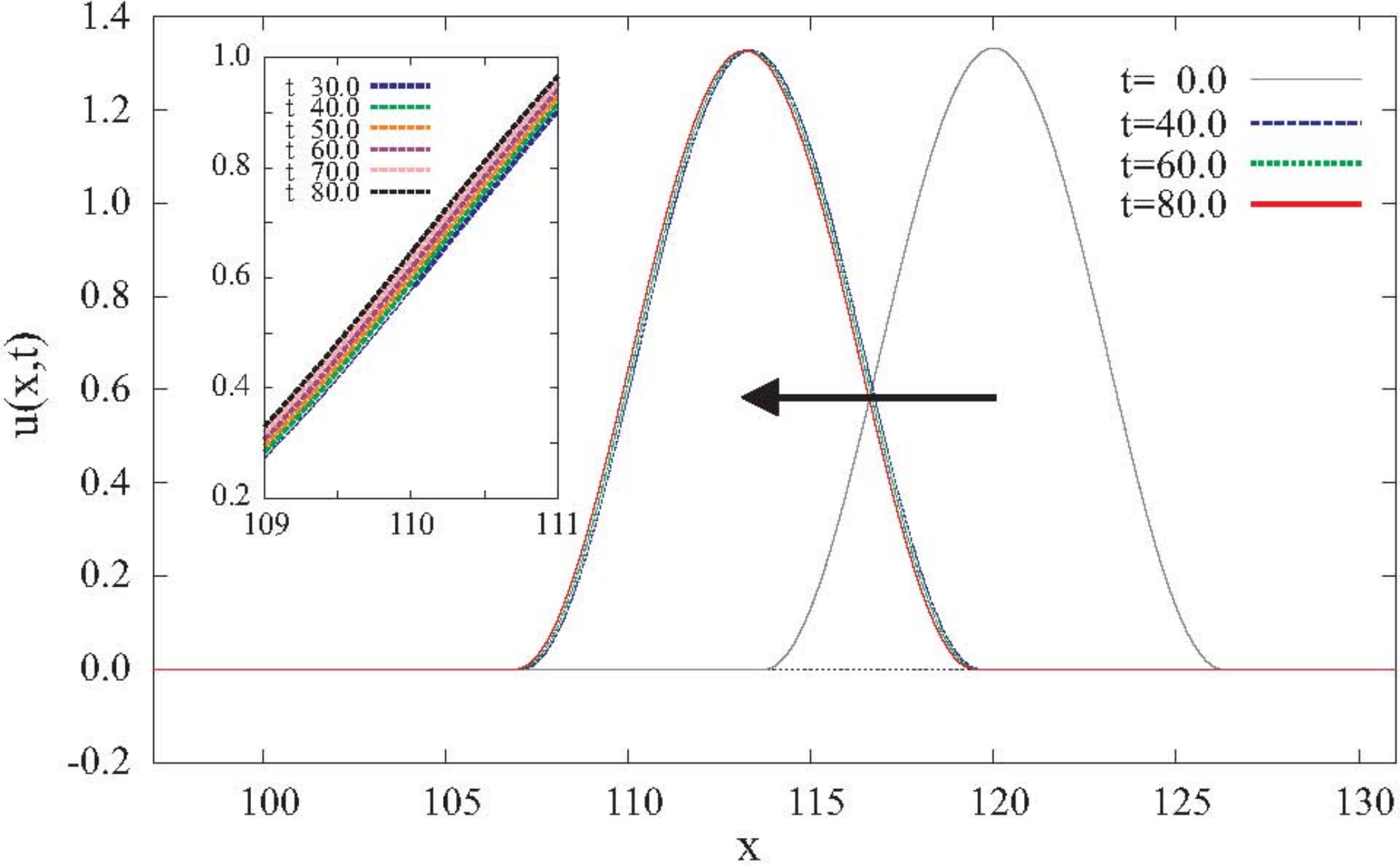}
   \caption{\label{coll_c1}(Color online)
   The first compacton ($c_1$=1) is ``at rest'' before the collision depicted in Fig.~\ref{coll_mov}. As shown in the inset (with the same axis labels), after the collision the centroid of this compacton changes position and the compacton moves slowly consistent with a small change in amplitude due to hyperviscosity.}
\end{figure}

%
%

\subsection{Study of compacton stability}

To illustrate a numerical study of a compacton stability problem, we apply the Pad\'e approximations discussed above to the case of the $K(2,2)$ compacton defined in Eq.~\eqref{eq:uc}. The numerical compactons propagate with the emission of forward and backward propagating radiation. In Fig.~\ref{stab_1}, we illustrate results for each numerical scheme by depicting the numerically-induced  radiation in the comoving frame of the compacton ($c_0=c$). Here we chose a snapshot at  $t$=175 after propagating the compacton in the absence of hyperviscosity ($\mu$=0) with a time step, $\Delta t$=0.002, and grid spacings, $\Delta x$=0.1, 0.05, and 0.02. We note that the amplitude of the radiation train is at least 7 orders of magnitude smaller than the amplitude of the compacton. Using the grid refining technique, we can show that indeed the radiation is a numerically-induced phenomenon. The noise is suppressed by reducing the grid spacing, $\Delta x$, indicating that the compacton~\eqref{eq:uc} is a stable solution of the $K(2,2)$  equation. 

For any numerical study of compacton stability and dynamical properties, it is important to reduce as much as possible the numerically-induced radiation.  It is desirable to minimize three characteristics of the radiation train: (i) the length of the radiation train, in order to avoid a wrap around of the solution as a result of the periodic boundary conditions constraint, (ii) the amplitude of the radiation train, which should be minimized  in order to better differentiate numerical artifacts from physics, and (iii) the amplitude at the leading edge of the radiation train, which seems related to the susceptibility of the numerical approximation to instabilities arising particularly in dynamical studies. Large amplitudes at the leading edge of the radiation train lead to the need for large values of the hyperviscosity parameter, $\mu$, in order to overcome these instabilities.

%
%

\begin{figure}[t!]
   \centering
   \includegraphics[width=0.9\columnwidth]{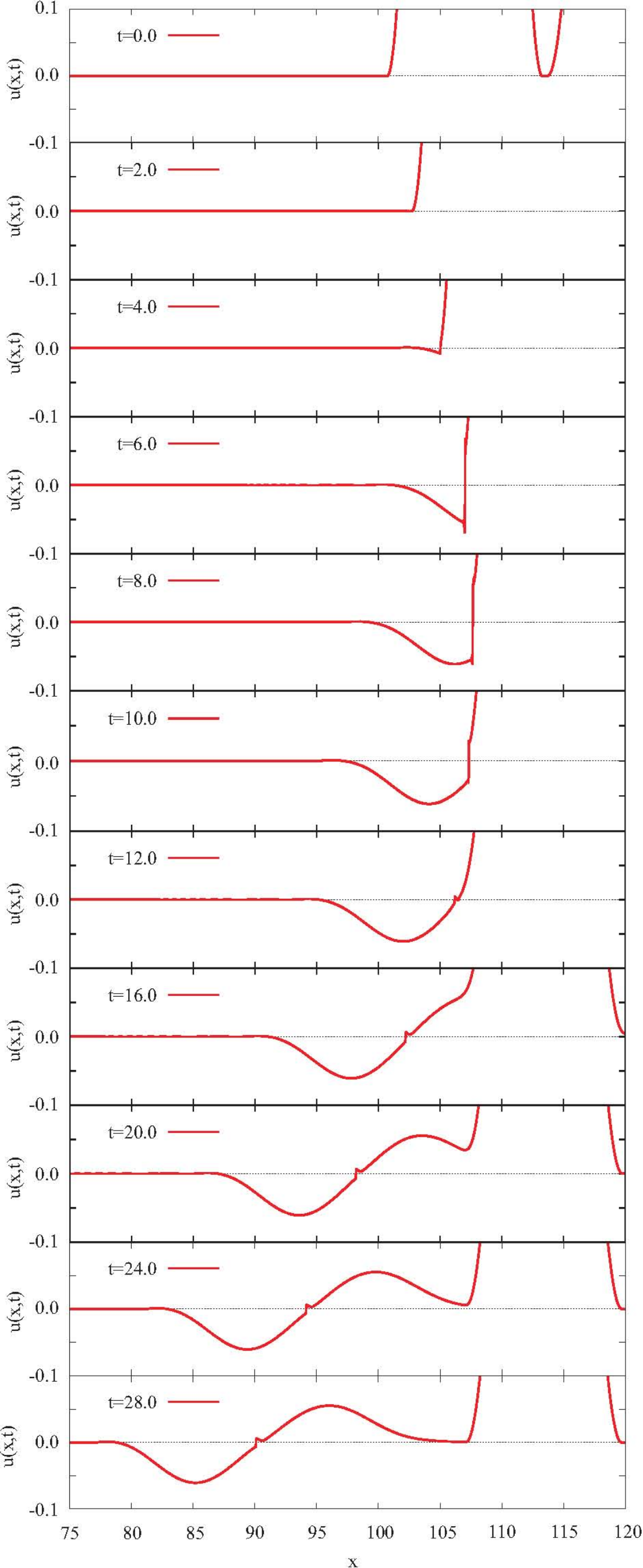}
   \caption{\label{coll_det}(Color online)
   Dynamics of the zero-mass``ripple'' with shock created as a result of the collision depicted in Fig.~\ref{coll_mov}. }
\end{figure}

%
%

To study the quality of our Pad\'e approximations, in Fig.~\ref{stab_2} we compare the radiation results at $t$=175 following the compacton propagation with $\Delta t$=0.002 and $\Delta x$=0.1 in the comoving frame of the compacton. We notice that the spatial extent of the radiation train is minimum in the optimal (4,4,4) scheme. 
Furthermore, the $K(p,p)$  equation, Eq.~\eqref{Kpp}, depends only on first- and third-order spatial derivatives. It appears, that at least for the $K(2,2)$  compacton, an improved first-order derivative approximation, i.e. the (6,4,4) scheme, leads to a shorter radiation train than in the case of the (4,4,6) scheme, which improves the quality of the third-order derivative~\cite{group_velocity}. However, the amplitude of the radiation train wave in the (4,4,6) scheme is comparable with the train amplitude in the (4,4,4) scheme, and smaller than the train amplitude in the (6,4,4) scheme, which indicates that it is important to improve the numerical approximation of the third-order spatial derivative in the $K(2,2)$ equation. The $K(2,2)$ equation does not feature a second-order spatial derivative, and the (4,6,4) scheme behaves as a tradeoff between the (6,4,4) and (4,4,6) schemes: the radiation train is shorter in the (4,6,4) scheme, but the amplitude of the train is comparable with that in the (6,4,4) scheme and larger than in the (4,4,6) scheme. Finally, with respect to the amplitude at the leading edge of the radiation train, the (4,4,6) scheme is the best and the (4,4,4) scheme is the worst and likely will require the largest hyperviscosity parameters in dynamical problems.

%
%

\begin{figure}[t!]
   \centering
   \includegraphics[width=0.9\columnwidth]{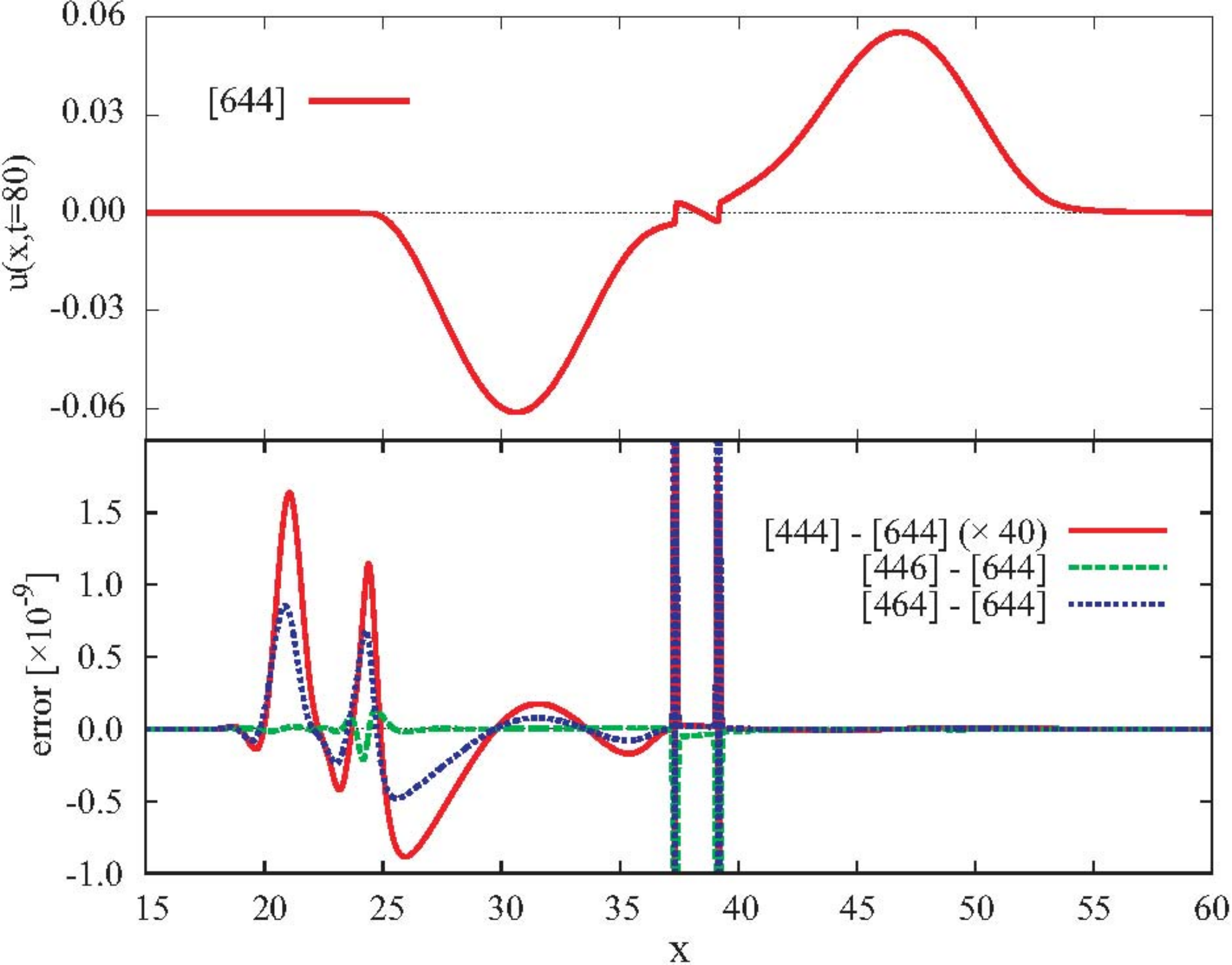}
   \caption{\label{coll_sh}(Color online)
   Comparison of the Pad\'e numerical schemes in the context of the ripple created as a result of the collision depicted in Fig.~\ref{coll_mov}. In the upper panel we illustrate the ripple calculated for $t$=80 using the (6,4,4) scheme, whereas in the bottom panel we illustrate the differences between results obtained using the other schemes and the (6,4,4) scheme. All simulations were performed using a hyperviscosity, $\mu=10^{-4}$.}
\end{figure}

\begin{figure}[t!]
   \centering
   \includegraphics[width=0.9\columnwidth]{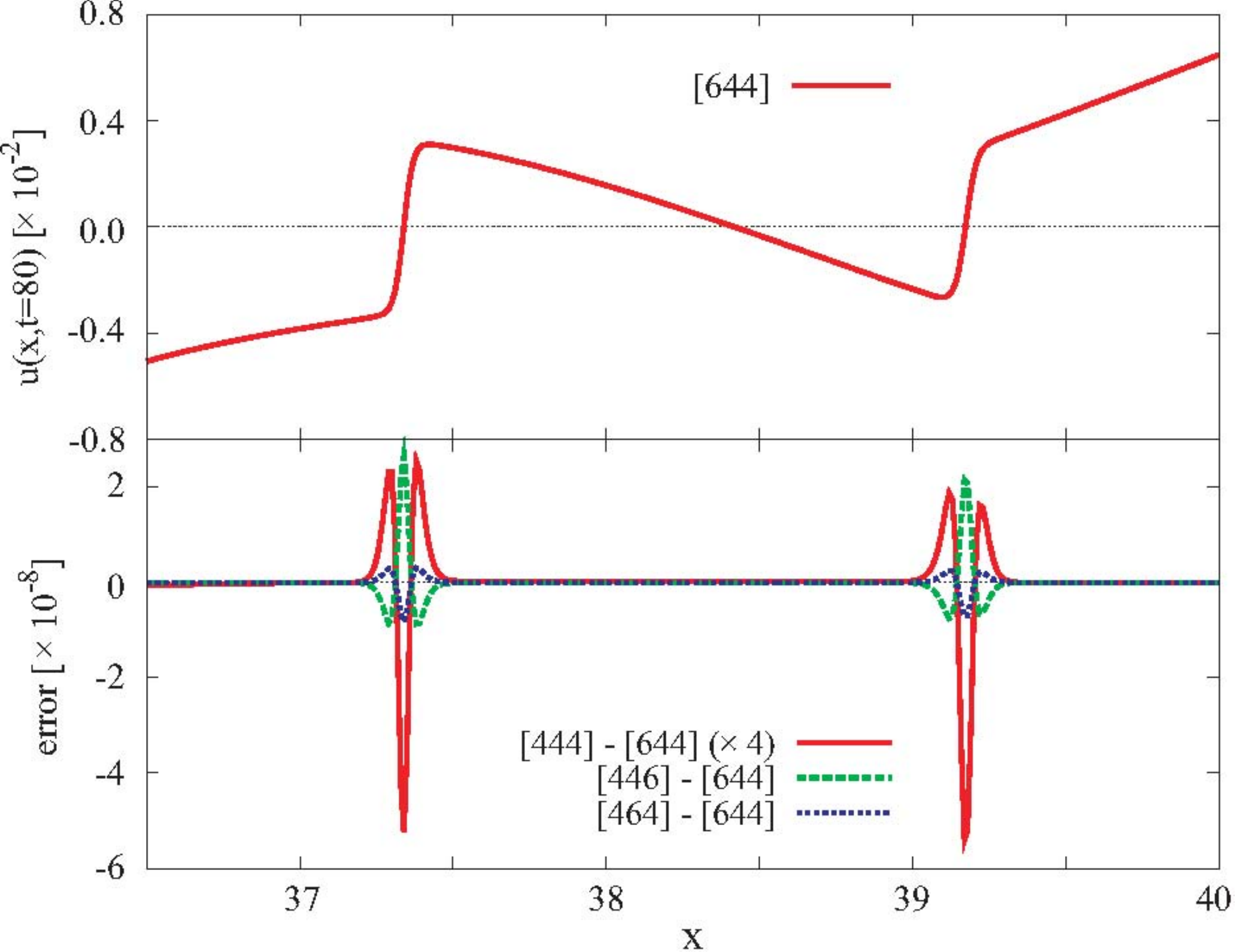}
   \caption{\label{coll_sh2}(Color online)
   Similar to Fig.~\ref{coll_sh}. Here we compare the numerical schemes in the context of the shock components observed in the ripple created as a result of the collision depicted in Fig.~\ref{coll_mov}.}
\end{figure}

%
%

\subsection{Study of compacton dynamics}

It is generally accepted that the RH compactons have the soliton property that after colliding with other compactons they reemerge with the same coherent shape. However, unlike in soliton collisions, the point where the compactons collide is marked by the creation of low-amplitude compacton-anticompacton pairs~\cite{RH93}. De Frutos \emph{et al.} showed~\cite{dF95}, and RV confirmed recently~\cite{RV07a}, that shocks are generated during compacton collisions. Shocks are also generated when arbitrary initial ``blobs'' decompose into a series of compactons~\cite{RV07a}. These shocks offer an ideal setting to compare numerical approximations  such as the Pad\'e approximants discussed here.


\subsubsection{Pairwise interaction of compactons}

In this scenario, we consider the collision between two compactons~\eqref{eq:uc} with velocities $c_1=1$ and $c_2=2$. In Fig.~\ref{coll_mov}, we depict a series of snapshots of this collision process. The compactons are  propagated in the comoving frame of reference of the first compacton, i.e. $c_0=c_1$, using the (6,4,4) scheme and a hyperviscosity, $\mu=10^{-4}$. The collision is shown to be inelastic, despite the fact that the compactons maintain their coherent shapes after the collision. The first compacton is ``at rest" before the collision occurs. After the collision, this compacton  emerges with the centroid located at a new spatial location, as illustrated in Fig.~\ref{coll_c1} . The inset in Fig.~\ref{coll_c1} shows the compacton moving slowly after collision, consistent with a small change in its amplitude.

The collision process  depicted in Fig.~\ref{coll_mov} gives rise to a zero-mass ripple with a shock when the ``ripple'' switches from negative to positive values (see Fig.~\ref{coll_det}). 
A  small change in the shock amplitude is noticed and is due to the presence of hyperviscosity.
These shock components were first noted in Ref.~\cite{dF95}, and were shown to be robust with respect to the numerical approximation in Ref.~\cite{RV07b}.
In Fig.~\ref{coll_sh} we compare results obtained with our four Pad\'e approximants schemes in the context of the ripple created as a result of the collision depicted in Fig.~\ref{coll_mov}. In the upper panel of Fig.~\ref{coll_sh} we show the result obtained using the (6,4,4) scheme at $t$=80, whereas in the bottom panel we illustrate the differences between results obtained using the other schemes and the (6,4,4) scheme. All simulations were performed using a hyperviscosity, $\mu=10^{-4}$. Similarly, in Fig.~\ref{coll_sh2} we compare our numerical schemes in the context of the shock components observed when the ripple switches from negative to positive values. 

Based on the results depicted in Figs.~\ref{coll_sh} and~\ref{coll_sh2}, we observe  that, independent of the numerical scheme, the largest errors occur at the end of the ripple in the direction of its propagation (see Fig.~\ref{coll_sh}), and the errors are very similar in the area of the two shock components (see Fig.~\ref{coll_sh2}). Furthemore we note that the results of schemes (6,4,4) and (4,4,4) are very similar. Given that based on our stability studies we concluded that scheme  (4,4,4) provides the most accurate set of Pad\'e approximants for the $K(2,2)$ problem, this leads us to use the (6,4,4) scheme as the reference for this comparison. Finally, the (6,4,4) results are closer to the (4,4,6) results than they are to the (4,6,4) results, which seems to indicate that for dynamical $K(2,2)$ problems (4,6,4) approximation scheme fares the worst, as the $K(2,2)$ equation does not depend on second-order spatial derivatives. 

%
%

\begin{figure}[t!]
   \centering
   \includegraphics[width=0.9\columnwidth]{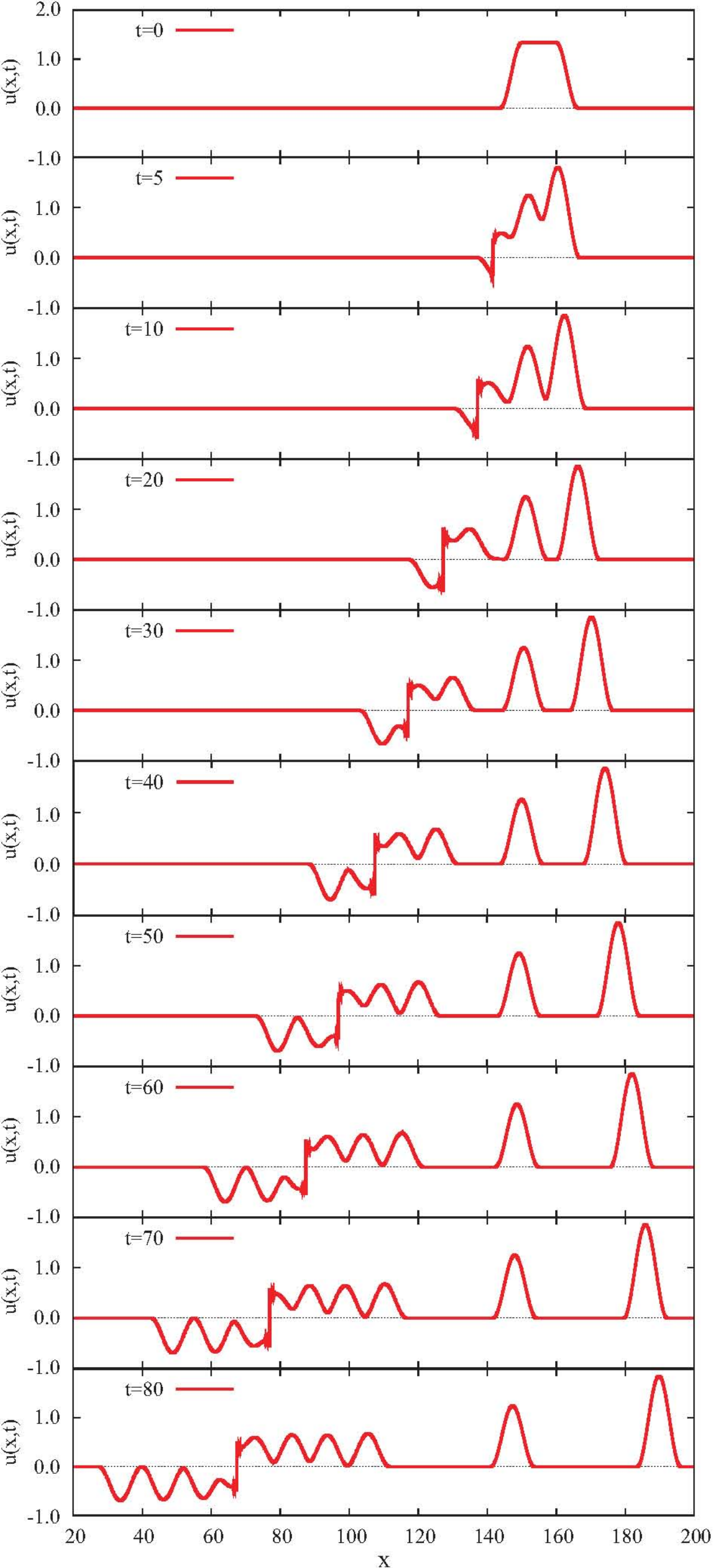}
   \caption{\label{decomp_mov}(Color online)
   Dynamics of a ``blob'' decomposition into two compactons and a ripple featuring a set of compacton-anticompacton pairs. Similar to the collision problem, the ripple has positive- and negative-value components, separated by a shock front. The simulation was performed using the (6,4,4) scheme and a  hyperviscosity, $\mu=10^{-4}$. }
\end{figure}

\begin{figure}[t!]
   \centering
   \includegraphics[width=0.9\columnwidth]{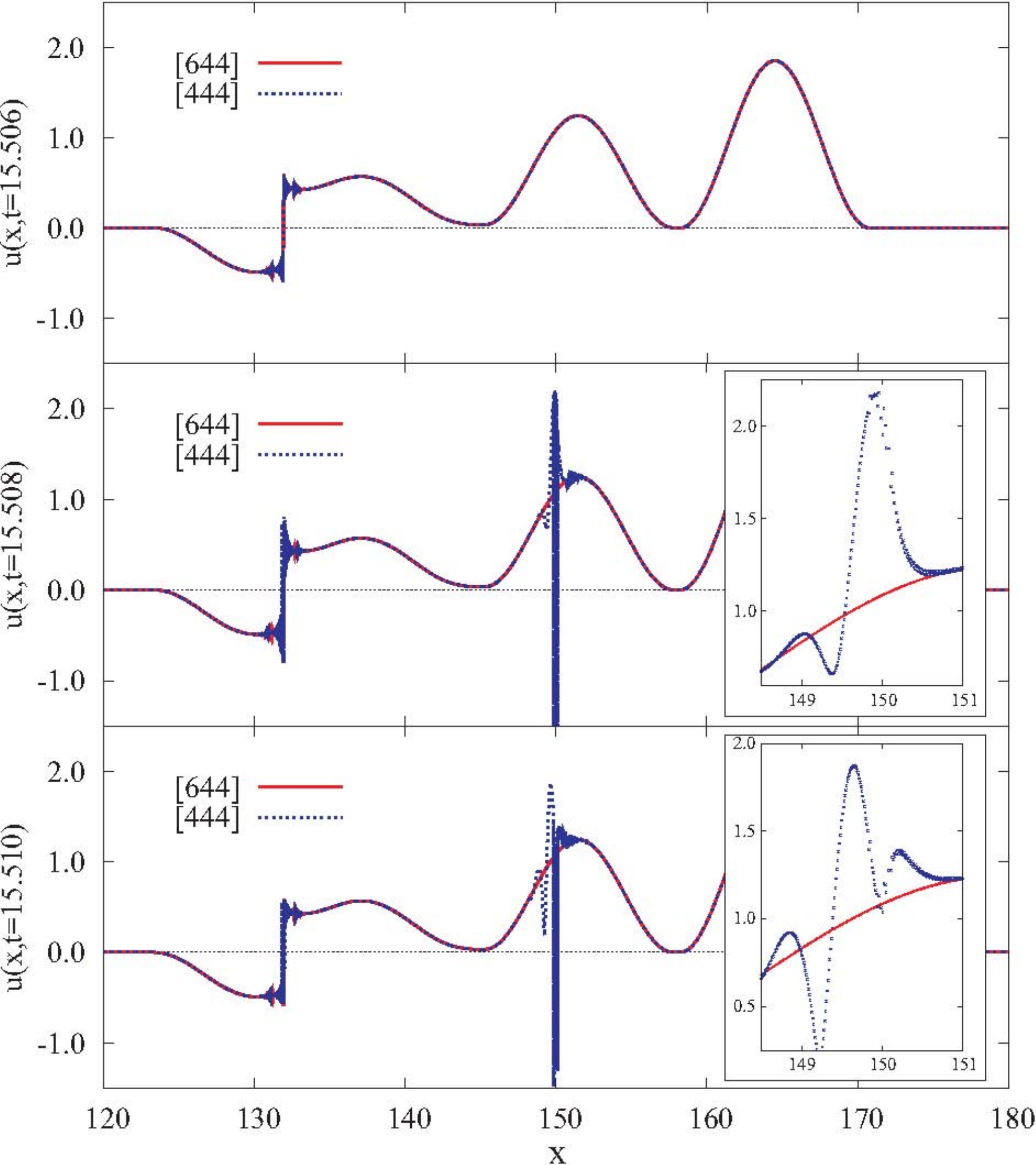}
   \caption{\label{decomp_444}(Color online)
   Autopsy of an instability: The simulation of the ``blob'' decomposition depicted in Fig.~\ref{decomp_mov}, becomes unstable and crashes when performed using the (4,4,4) scheme with a hyperviscosity, $\mu=10^{-4}$. Here we illustrate the last three steps in the simulation by comparing the results obtained using the (6,4,4) and (4,4,4) schemes. The insets show a 
   magnified region around $x=150$ for clarity. }
\end{figure}

\begin{figure}[t!]
   \centering
   \includegraphics[width=0.9\columnwidth]{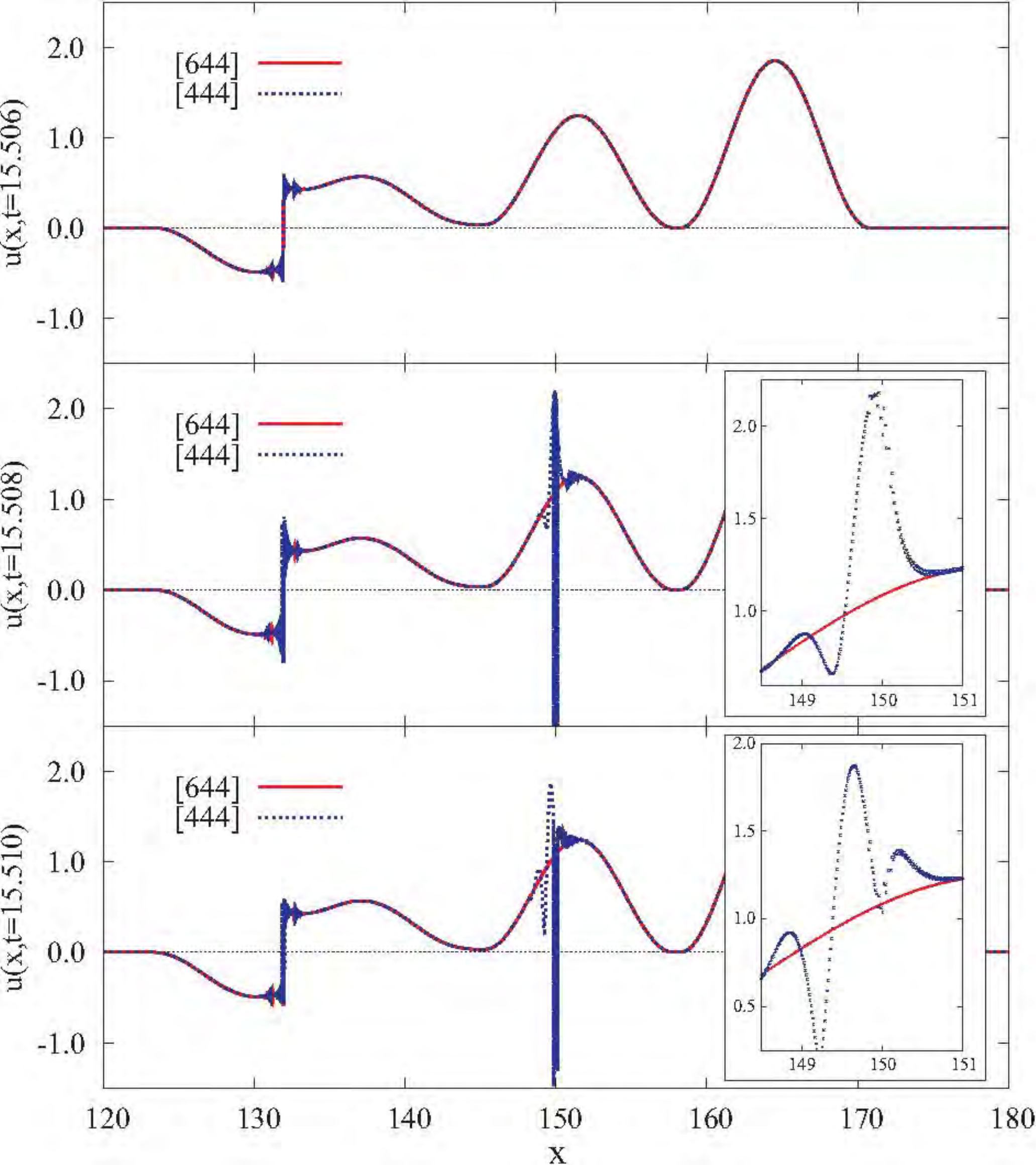}
   \caption{\label{decomp_444_eps}(Color online)
   Smoothing effect of the hyperviscosity: The simulation of the ``blob'' decomposition performed using the (4,4,4) scheme becomes unstable for a hyperviscosity $\mu=10^{-4}$. The instability cannot be removed by reducing the grid step (not shown). Instead one must increase the value of the hyperviscosity, which results in a smoothing of the noise at the shock front.}
\end{figure}

\begin{figure}[t!]
   \centering
   \includegraphics[width=0.9\columnwidth]{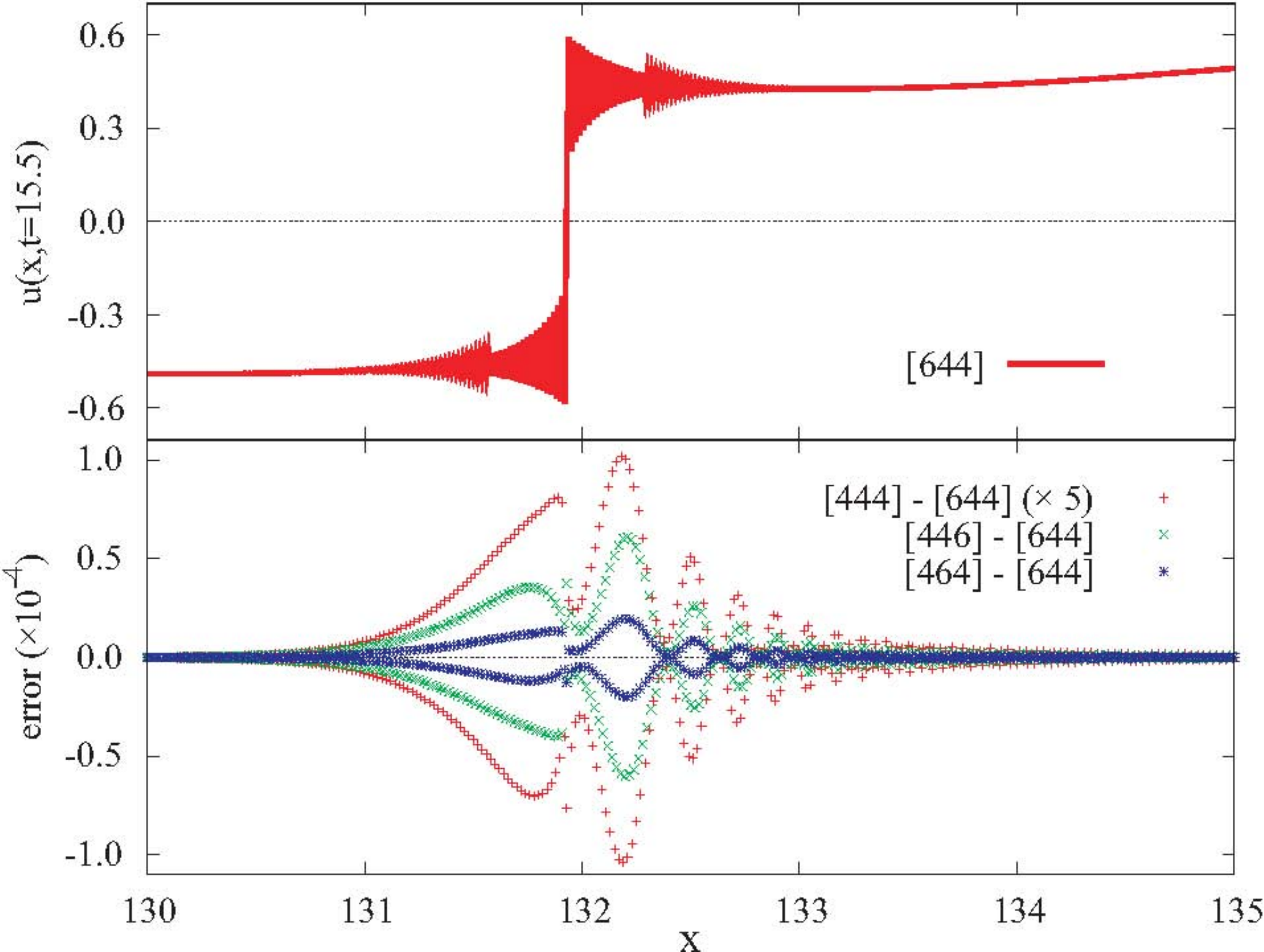}
   \caption{\label{decomp_31}(Color online)
   Comparison of the numerical schemes discussed in the context of the shock front formed when the ripple switches from negative to positive values. In the upper panel we depict the result at $t$=15.5 obtained using the (6,4,4) scheme, whereas in the bottom panel we illustrate the differences between results obtained using the other schemes and the (6,4,4) scheme. All simulations were performed using a hyperviscosity, $\mu=10^{-4}$. Shortly after $t=15.5$ the simulation performed using the (4,4,4) scheme becomes unstable.}
\end{figure}

\begin{figure}[t!]
   \centering
   \includegraphics[width=0.9\columnwidth]{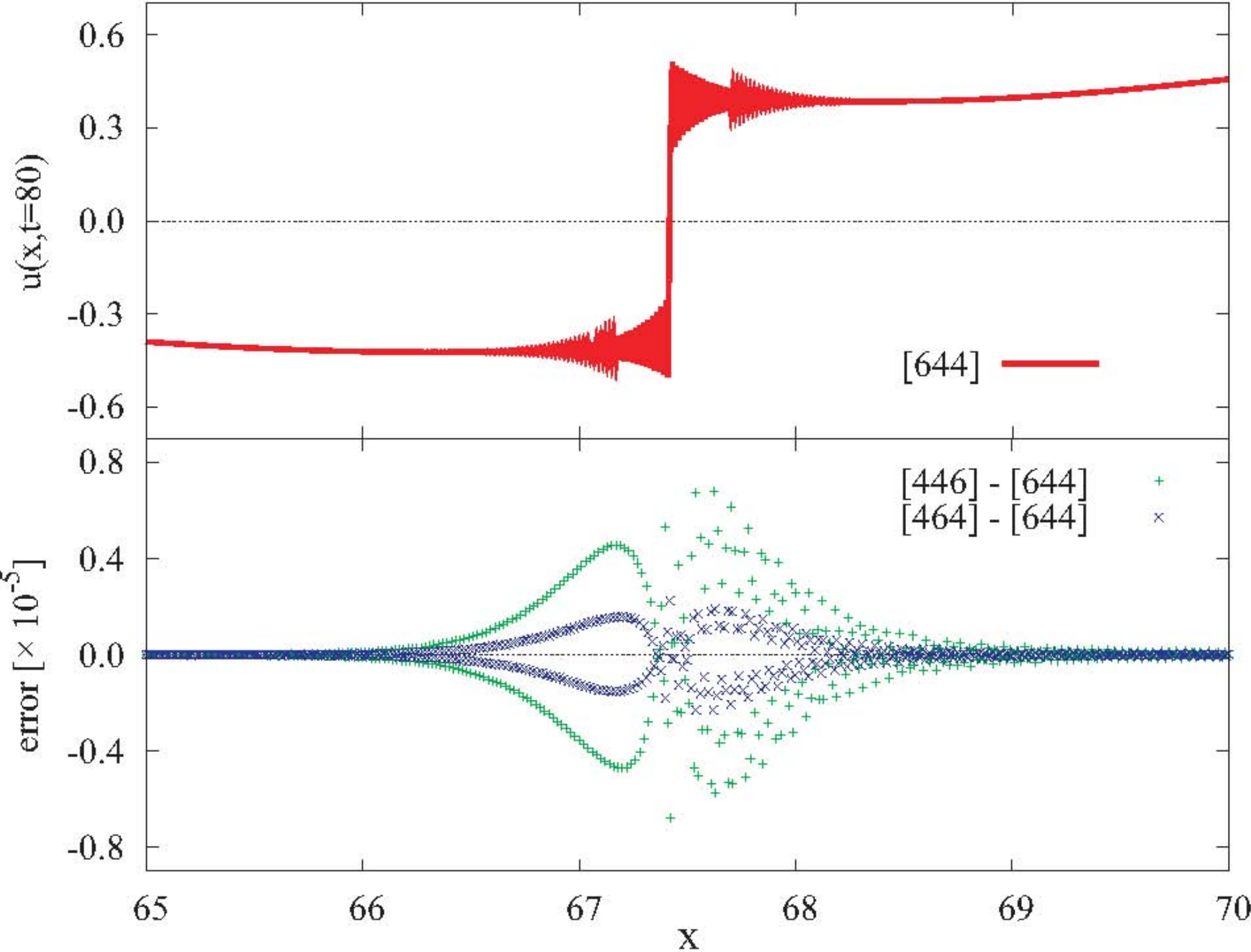}
   \caption{\label{decomp_160}(Color online)
   Similar to Fig.~\ref{decomp_31}. Here, results are presented for $t$=80.
   The simulation performed using the (4,4,4) scheme became unstable shortly after $t=15.5$ and is not represented here.}
\end{figure}

%
%

\subsubsection{Dynamics with arbitrary initial conditions}

Following RV~\cite{RV07a}, we consider the time evolution of a ``blob'' given 
\begin{equation}
    u(x,0) = 
    \Biggl \lbrace
    \begin{array}{ll}
    \frac{4c}{3} \cos^2 \frac{x-150}{4} ,
    & \mathrm{for} \  150 - 2\pi \leq x \leq 150 \>, \\
    \frac{4c}{3} ,
    & \mathrm{for} \  150 \leq 160 \>, \\
    \frac{4c}{3} \cos^2 \frac{x-160}{4} ,
    & \mathrm{for} \  160 -\leq x \leq 160 + 2\pi \>. \\
    \end{array}
\label{blob}
\end{equation}
In Fig.~\ref{decomp_mov} we illustrate the dynamics of this blob decomposition, as calculated using the (6,4,4) scheme and a  hyperviscosity, $\mu=10^{-4}$, and show that the blob evolves into two compactons and a ripple featuring a set of compacton-anticompacton pairs. Similar to the collision problem, the ripple has positive- and negative-value components separated by a shock. 

As surmised following our compacton stability study, the radiation train corresponding to the (4,4,4) scheme has a higher amplitude at the leading edge, which makes this scheme more susceptible to instabilities than the other three schemes. This undesirable feature of the (4,4,4) scheme is illustrated in the case of this blob decomposition. In Fig.~\ref{decomp_444}, we depict the last three steps in the simulation and compare results obtained using the (6,4,4) and (4,4,4) schemes. These results show how the (4,4,4) simulation of the ``blob'' decomposition  becomes unstable and crashes corresponding to a hypeviscosity, $\mu=10^{-4}$, whereas the (6,4,4) simulation does not. The (4,4,4) simulation becomes stable if we increase the hyperviscosity (see Fig.~\ref{decomp_444_eps}), but the instability cannot be removed by reducing the grid spacing (not shown). 

With the exception of the instability developed in the (4,4,4) scheme, the comparison of the four Pad\'e schemes reveals a situation very similar to the case of the pairwise compacton collision and results are illustrated in Fig.~\ref{decomp_31} at $t$=15.5 [just prior to scheme (4,4,4) becoming unstable] and Fig.~\ref{decomp_160}  at $t$=60. By comparing the four approximation schemes in the context of the shock front formed when the ripple switches from positive to negative values, we find that results obtained using schemes (4,4,4) and (6,4,4) are very close to each other and the (6,4,4) results are closer to the (4,4,6) results than they are to the (4,6,4) results. Unfortunately, the (4,4,4) scheme requires a larger hyperviscosity to smooth out the noise at the shock front. Our study suggests that the (6,4,4) scheme is the best alternative to scheme (4,4,4).

%
%

\section{Conclusions}
\label{sec:concl}

To summarize, in this paper we presented a systematic approach to calculating higher-order derivatives of smooth functions on a uniform grid  using Pad\'e approximants. We illustrated this approach by deriving higher-order approximations using traditional second-order finite-differences formulas as our starting point. We proposed four special cases of the fourth-order Pad\'e approximants and employed these schemes to study the stability and dynamical properties of  $K(p,p)$  Rosenau-Hyman compactons. This study was designed to establish the baseline for future studies of the stability and dynamical properties of $L(r,s)$ compactons, Eq.~\eqref{eq:Llp},  that unlike the RH compactons are derivable from a Lagrangian ansatz. The $L(r,s)$ compactons feature higher-order nonlinearities and terms with mixed-derivatives that are not present in the $K(p,p)$ equations, and hence the numerical analysis of their properties is expected to be considerably more difficult.  Finally, we intend to apply these schemes to the study of PT-symmetric compactons ~\cite{BCKMS09}. 

Based on our compacton stability study, we conclude that none of our four Pad\'e approximations appears as a clear winner in the context of the three minimization criteria for an optimal numerical discretization of the compacton problem:  length and amplitude of the radiation train, and amplitude at the leading edge of the radiation train. The (4,4,4) scheme features the shortest and smallest amplitude of the radiation train, but exhibits the largest amplitude at the leading edge of the train. The (4,4,6) scheme has the smallest amplitude at the leading edge of the train and the amplitude of the train is comparable otherwise with that in the (4,4,4) scheme, but the length of the radiation train in the (4,4,6) scheme is the largest of the four approximations considered here. Hence, the (4,4,6) scheme will probably be most useful in studies of the dynamical properties of the system and deal best with shock-type problems, but will require the largest model space (largest value of $L$) to avoid the undesired wrap around effect of the solution due to the periodic boundary conditions constraint, which in turn will make grid refining studies difficult in the (4,4,6) scheme, because of physical constraints in allocatable memory and CPU wall time. 

The best compromise may be provided by the (6,4,4) approximation scheme. Our dynamical simulations indicate that the results obtained using this method closely resemble those obtained using the (4,4,4) scheme, and the (6,4,4) scheme is less susceptible to instabilities, at least in the particular cases discussed here, than the (4,4,4) scheme.

\begin{acknowledgments} 
   This work was performed in part under the auspices of the United States Department of Energy.  
   The authors gratefully acknowledge useful conversations with C. Mihaila, F. Rus and F.R. Villatoro. 
   B. Mihaila and F. Cooper would like to thank the Santa Fe Institute for its hospitality during the completion of this work.
\end{acknowledgments}

\vfill


%
%


\appendix

\section{}
\label{app:A}

In this appendix we discuss the derivation of fourth-order accurate approximations for the first- and second-order derivatives of a smooth function.

To derive a fourth-order accurate approximation of the first-order derivative, we begin by introducing the operator
\begin{align}
   &
   \mathcal{\tilde A}_1(E) \, \um
   =
   \frac{1}{\alpha \del} \,
   \Bigl [ \bigl ( E^2 - E^{-2} \bigr )
            + \beta \bigl ( E - E^{-1} \bigr )
   \Bigr ] \, \um
   \>,
\end{align}
and ask that the following relation is fulfilled:
\begin{align}
   \uI
   =
   \mathcal{\tilde A}_1 (E) \, \um
   + \mathcal{O}(\del^4)
   \>.
\end{align}
We have
\begin{align}
   &
   \mathcal{\tilde A}_1(E) \, \um
   =
   \frac{1}{\alpha} \,
   \Bigl \{
   4 \, \uI
   + \uIII \frac{8 \del^2}{3}
   + \uV \frac{8 \del^4}{15}
   \notag \\ & \qquad
   + \uVII \frac{16 \del^6}{315}
   + \cdots
   + \beta \,
   \Bigl [
   2 \, \uI
   + \uIII \frac{\del^2}{3}
   \notag \\ & \qquad
   + \uV \frac{\del^4}{60}
   + \uVII \frac{\del^6}{2520}
   + \cdots
   \Bigr ]
   \Bigr \}
   \>.
\end{align}
To satisfy the requirement of a fourth-order accurate approximation for the first-order derivative $\uI$, we solve the system of equations
\begin{align}
   \alpha - 2 \beta = & \ 4
   \>,
   \\
   \beta = & \, - 8
   \>,
\end{align}
and obtain the solution
\begin{align}
   \alpha = -12 \>, \quad
   \beta = -8 \>.
\end{align}
This gives
\begin{align}
   \uI
   = & \
   \mathcal{\tilde A}_1(E) \, \um
   + \uV \frac{\del^4}{30}
   + \mathcal{O}(\del^6)
   \>,
\end{align}
with
\begin{align}
   \mathcal{\tilde A}_1(E) & \ =
   - \,
   \frac{1}{12 \del} \,
   \Bigl [
      E^2 - 8 E + 8 E^{-1} - E^{-2}
   \Bigr ]
   \>.
\end{align}

Similarly, to derive a fourth-order accurate approximation for the second-order derivative we introduce the operator
\begin{align}
   &
   \mathcal{\tilde B}_1(E) \, \um
   =
   \frac{1}{\alpha \del^2} \,
   \Bigl [ \bigl ( E^2 + E^{-2} \bigr )
            + \beta \bigl ( E + E^{-1} \bigr )
            + \gamma
   \Bigr ] \, \um
   \>,
\end{align}
and seek $\alpha$, $\beta$, and $\gamma$ such that
\begin{align}
   \uII
   =
   \mathcal{\tilde B}_1 (E) \, \um
   + \mathcal{O}(\del^4)
   \>.
\end{align}
We have
\begin{align}
   &
   \mathcal{\tilde B}_1(E) \, \um
   =
   \frac{1}{\alpha \del^2} \,
   \Bigl \{
   2 \um
   + 4 \uII \del^2
   + \uIV \frac{4 \del^4}{3}
   + \cdots
   \notag \\ & \
   + \beta \,
   \Bigl [
   2 \um
   + \uII \del^2
   + \uIV \frac{\del^4}{12}
   + \cdots
   \Bigr ]
   + \gamma \, \um
   \Bigr \}
   \>,
\end{align}
which gives the system of equations
\begin{align}
   2 \beta + \gamma = & \ - 2
   \>,
   \\
   \alpha - \beta = & \ 4
   \>,
   \\
   3 \alpha - \beta = & \ 0
   \>,
\end{align}
with the solution
\begin{align}
   \alpha = -2 \>, \quad
   \beta = -6 \>, \quad
   \gamma = 10 \>.
\end{align}
Hence, we obtain
\begin{align}
   \mathcal{\tilde B}_1(E) & \ =
   - \, \frac{1}{2 \del^2} \,
   \Bigl [
      E^2 - 6 E + 10 - 6  E^{-1} + E^{-2}
   \Bigr ]
   \>.
\end{align}
and
\begin{align}
   \uII
   = & \
   \mathcal{\tilde B}_1(E) \, \um
   + \uVI \frac{5}{12} \, \del^4
   + \mathcal{O}(\del^6)
   \>.
\end{align}


\begin{thebibliography}{99}
\bibitem{group_velocity}
A better analysis of the quality of these numerical schemes can be done using the analytical approximation of the group velocity of the radiation train discussed in Ref.~\onlinecite{RV07b} and will be carried out in the future.
\end{thebibliography}
\end{document}